\documentclass[a4paper,fleqn,usenatbib]{mnras}
\usepackage{mathptmx}
\usepackage[T1]{fontenc}
\usepackage{ae,aecompl}
\usepackage{times}
\usepackage{amssymb}
\usepackage{amsmath}
\usepackage{upgreek}
\usepackage{color}
\usepackage{graphicx}
\usepackage{pdflscape}

\title[JPS compact source catalogue]{The JCMT Plane Survey: First complete data release --- emission maps and compact source catalogue}
\author[D. J. Eden et al.]{D. J. Eden,$^{1}$\thanks{E-mail: D.J.Eden@ljmu.ac.uk} T.J.T. Moore,$^{1}$ R. Plume,$^{2}$ J.\,S.\,Urquhart,$^{3}$ M.A. Thompson,$^{4}$ \newauthor H. Parsons,$^{5}$ J.T. Dempsey,$^{5}$ A.J. Rigby,$^{6}$ L.K. Morgan,$^{1}$ H.S. Thomas,$^{5}$ D. Berry,$^{5}$ \newauthor J. Buckle,$^{7,8}$ C.M. Brunt,$^{9}$ H.M. Butner,$^{10}$ D. Carretero,$^{8}$ A. Chrysostomou,$^{11}$ \newauthor M.J. Currie,$^{5}$ H.M. deVilliers,$^{4}$ M. Fich,$^{12}$ A.G. Gibb,$^{13}$ M.G. Hoare,$^{14}$ \newauthor T. Jenness,$^{15}$ G. Manser,$^{4}$ J.C. Mottram,$^{16}$ C. Natario,$^{4}$ F. Olguin,$^{14}$ N. Peretto,$^{6}$ \newauthor M. Pestalozzi,$^{17}$ D. Polychroni,$^{18}$ R.O. Redman,$^{5}$ C. Salji,$^{8}$ L.J. Summers,$^{9}$ K. Tahani,$^{2}$ \newauthor A. Traficante,$^{17}$ J. diFrancesco,$^{19}$ A. Evans,$^{20}$ G.A. Fuller,$^{21}$ D. Johnstone,$^{19,22}$ \newauthor G. Joncas,$^{23}$ S.N. Longmore,$^{1}$ P.G. Martin,$^{24}$ J.S. Richer,$^{8}$ B. Weferling,$^{5}$ G.J. White,$^{25,26}$ \newauthor and M. Zhu$^{27}$\\
Affiliations are listed at the end of the paper}

\date{Accepted XXX. Received YYY; in original form ZZZ}

\pubyear{2016}

\begin{document}
\label{firstpage}
\pagerange{\pageref{firstpage}--\pageref{lastpage}}
\maketitle

\begin{abstract}

We present the first data release of the James Clerk Maxwell Telescope (JCMT) Plane Survey (JPS), the JPS Public Release 1 (JPSPR1). JPS is an 850-$\upmu$m continuum survey of six fields in the northern inner Galactic Plane in a longitude range of $\ell$\,=\,7$\degr$--63$\degr$, made with the Sub-millimetre Common-User Bolometer Array 2 (SCUBA-2). This first data release consists of emission maps of the six JPS regions with an average pixel-to-pixel noise of 7.19\,mJy\,beam$^{-1}$, when smoothed over the beam, and a compact-source catalogue containing 7,813 sources. The 95 per cent completeness limits of the catalogue are estimated at 0.04\,Jy\,beam$^{-1}$ and 0.3\,Jy for the peak and integrated flux densities, respectively. The emission contained in the compact-source catalogue is 42\,$\pm$\,5 per cent of the total and, apart from the large-scale (greater than 8\,arcmin) emission, there is excellent correspondence with features in the 500-$\upmu$m \emph{Herschel} maps. We find that, with two-dimensional matching, 98\,$\pm$\,2 per cent of sources within the fields centred at $\ell$\,=\,20$\degr$, 30$\degr$, 40$\degr$ and 50$\degr$ are associated with molecular clouds, with 91\,$\pm$\,3 per cent of the $\ell$\,=\,30$\degr$ and $40\degr$ sources associated with dense molecular clumps. Matching the JPS catalogue to {\em Herschel} 70-$\upmu$m sources, we find that 38\,$\pm$\,1 per cent of sources show evidence of ongoing star formation. The images and catalogue will be a valuable resource for studies of star formation in the Galaxy and the role of environment and spiral arms in the star formation process.

\end{abstract}


\section{Introduction}

The production of a predictive model for star formation requires a number of key questions to be answered. It is crucial to determine what mechanisms control the star-formation rate (SFR) and efficiency (SFE), on what scales they operate, and whether any of these mechanisms also cause variations in the stellar initial mass function (IMF). Any predictive model will also need to factor in the influence of the environment in which the star formation is occurring.

Until recent advancements in telescope facilities and instrumentation, studies trying to answer these questions focused on individual star-forming regions, but now survey-driven research has started to make progress in addressing the issues \citep[e.g.][]{Schuller09,Molinari10}.

The first step of the process, after or during the formation of molecular clouds, is the formation of the dense clumps and cores within which stars form. Studies such as \citet{Eden12,Eden13} and \citet{Battisti14}, using the molecular clouds detected in the Galactic Ring Survey (GRS; \citealt{Jackson06}) and dense clumps from the Bolocam Galactic Plane Survey (BGPS; \citealt{Aguirre11}), have made progress in examining the efficiency of this step using relatively small samples.  \cite{Eden12,Eden13} found no significant variation on kiloparsec scales in the clump-formation efficiency (CFE, also known as the dense-gas mass fraction, or DGMF\footnote{The CFE or DGMF is calculated by dividing the mass in dense structures by the molecular gas mass, traced by CO.}). These studies have revealed that the CFE is relatively constant for spiral-arm and inter-arm regions and a near-constant CFE value of $\sim$\,8 per cent \citep[i.e.][]{NguyenLuong11,Battisti14,Barnes16}.

To measure the stellar IMF and the SFE and, particularly, to detect variations in them, studies of large samples of young stellar objects (YSOs) are required. The IMF can be inferred from measurements of the luminosity function (LF) of the YSOs and an analogue of the SFE can be obtained from the ratio of IR luminosity to the cloud/clump mass reservoir (assuming the IR-bright timescale is short).  Galactic-scale samples of YSOs from the Red MSX Source survey (RMS; \citealt{Lumsden13}) were examined by \citet{Moore12}, who found that $\sim$70 per cent of the increase in SFR density in the spiral arms is due to source crowding, rather than a physical effect caused by the spiral arms themselves.  Much of the remainder could be ascribed to individual extreme sources.  For instance, they suggested that a steeper luminosity function in the W49A high-mass star-forming region may be responsible for a large increase in SFE in a section of the Perseus spiral arm. 

Based on the observed similarity between them, the development of the mass distribution of dense clumps (or clump mass function; CMF) may be the stage at which the slope of the IMF is set, with a constant conversion efficiency between the two \citep[e.g.][]{Beltran06}. Simulations have found that the lognormal density fluctuations in a turbulent medium set the CMF, and consequently the IMF \citep{Hennebelle08}. \citet{Hopkins12} added to this, finding that the mass function of bound objects is set by the smallest scale on which they are self-gravitating, with further work indicating that the slope and turnover mass can also be replicated \citep{Guszejnov15}. Any differences in the CMF may therefore hint at variations in the IMF, and a different mechanism of star formation, but clustered clumps have the same mass-function slope as those that form in an isolated environment \citep{Beuret17}, indicating that the CMF may be invariant.

The environment in which star formation occurs must be relevant, as the initial conditions for star formation in the Outer Galaxy differ significantly from those in the Inner Galaxy, with decreased metallicity, thermal and turbulent pressure, radiation field, and spiral-arm strength, to name a few. \citet{Roman-Duval09} and \citet{Moore12} found evidence that the mean mass of molecular clouds decreases with increasing Galactocentric radius, although it is not yet clear that this trend is not a statistical or selection effect. Other trends with Galactocentric radius have been found, with \citet*{Koda16} finding that the molecular gas mass fraction (amount of atomic gas converted to molecular gas) decreases rapidly with increasing Galactocentric radius. Similarly, using $^{13}$CO as a tracer of dense molecular gas, the fraction of molecular gas ($^{12}$CO + $^{13}$CO) converted into dense molecular gas also decreases with Galactocentric radius beyond $\sim$4\,kpc \citep{Roman-Duval16}. The possibility that these trends could be affecting star formation in the Outer Galaxy is supported by the recent finding by \citet{Ragan16} that the fraction of clumps containing a tracer of star formation decreases with Galactocentric radius. The reasons for this behaviour are unclear, however, as the CFE has been found to be constant across Galactocentric radius \citep{Eden13,Battisti14}. In the central regions of galaxies, \citet{James16} found that in external galaxies, the areas swept out by bars have suppressed star-formation rates, with the central molecular zone of our Galaxy having suppressed star-formation rates \citep{Longmore13}.

Other evidence from large-scale surveys suggests that it is within individual clumps and molecular clouds that the dominant variations in star-forming conditions occur, with values of CFE and SFE found to be log-normal from clump-to-clump and cloud-to-cloud \citep{Eden12,Eden15}. There is also evidence that massive clumps that are forming high-mass stars are more compact and have more strongly peaked surface-brightness distributions in the sub-millimetre continuum than those that are not \citep{Csengeri14,Urquhart14}.  Also, the structure of a high-mass star-forming clump appears to be set before the onset of star formation, and changes little as the embedded object evolves towards the main sequence.

\subsection{Complementary Galactic Plane Surveys}

These results are a consequence of the combination of multiple Galactic-Plane surveys tracing all of the important stages of the star-formation process from the molecular gas to YSOs and other signposts of young stars, from radio wavelengths to the infrared. 

The molecular gas in the JPS region is detected by observations in different rotational transition lines of CO isotopologues. Using the $J = 1-0$ transition, the FUGIN (FOREST Ultra-wide Galactic Plane survey In Nobeyama) survey will be observing the Northern Galactic Plane in the isotopologues $^{12}$CO, $^{13}$CO and C$^{18}$O with the Nobeyama 45-m Radio Telescope \citep{Minamidani16}, matching the angular resolution of the JPS. The GRS has also observed the inner Galactic Plane in $^{13}$CO $J = 2-1$, and has been used to produce a catalogue of molecular clouds, complete with distances and masses \citep{Rathborne09,Roman-Duval09,Roman-Duval10}. The $J = 2-1$ transition is covered by the SEDIGISM survey in the $^{13}$CO and C$^{18}$O isotopologues (Schuller et al., in press), extending to $\ell$\,=\,17$\degr$, covering the southern regions not covered by the GRS. The COHRS \citep*{Dempsey13a} and CHIMPS \citep{Rigby16} surveys cover all three isotopologues in the $J = 3-2$ transition at the same angular resolution as the JPS, but do not provide complete longitude or latitude coverage.

Star-formation tracers occur across the electromagnetic spectrum, with YSOs identified in the infrared by the RMS survey's colour-selected Galaxy-wide samples \citep{Lumsden13}. The radio-continuum observations of the CORNISH survey at 5\,GHz using the Very Large Array \citep{Hoare12} detected compact \ion{H}{ii} regions in the northern inner Galactic Plane \citep{Urquhart13}. The Methanol Multi-Beam Survey (MMB; \citealt{Green12}) has surveyed the entire Plane at 6.7\,GHz, revealing a comprehensive catalogue of methanol masers. The presence of such emission is an indicator of early high-mass star formation, with 99 per cent of masers found to be associated with compact submillimetre continuum emission \citep{Urquhart15}.

Continuum emission at submillimetre and far-infrared wavelengths traces the cold dust which is assumed to be well mixed with the gas. Several surveys have used these wavelengths to trace regions of current and incipient star formation in the form of dense clumps. The BGPS surveyed the northern Galactic Plane at 1.1-mm \citep{Aguirre11,Ginsburg13} with the ATLASGAL survey covering the majority of the JPS range at 870\,$\upmu$m \citep{Schuller09}. The entire Galactic Plane has been observed using the \emph{Herschel Space Observatory} at 70, 160, 250, 350 and 500\,$\upmu$m \citep{Molinari10a,Molinari10} with the images and single-band catalogues for the inner Plane published recently \citep{Molinari16,Molinari16a}. Hi-GAL surveys the Galactic Plane at resolutions of 6, 12, 18, 24 and 35 arcsec for the five wavebands, with minimum sensitivities of 0.5, 3, 5.5, 7 and 7\,Jy, respectively, across the JPS longitudes.

\subsection{JCMT Plane Survey}

Adding to the surveys in the continuum, we present here the first complete public release of data, JPS Public Release 1 (JPSDR1), and the compact-source catalogue extracted from the James Clerk Maxwell Telescope (JCMT) Plane Survey (JPS). This survey is part of the JCMT Legacy Survey programme (JLS; \citealt{Chrysostomou10}), a series of surveys studying star formation across the Universe from local Galactic studies to high-redshift galaxies \citep[e.g.][]{Wilson12,Kirk16,Geach17}.

JPS is a targeted, yet unbiased, survey of the inner Galactic Plane in the longitude range 7$\degr$\,<\,$\ell$\,<\,63$\degr$ using the wide-field sub-mm-band bolometer camera, the Sub-millimetre Common-User Bolometer Array 2 (SCUBA-2; \citealt{Holland13}) at 850\,$\upmu$m with an angular resolution of 14.5 arcsec. The target rms sensitivity was 10\,mJy\,beam$^{-1}$, when smoothed over the beam, and the achieved rms values are significantly better (Table~\ref{rmsvalues}). The unsmoothed pixel-to-pixel rms values are used and referred to for the rest of this paper. The survey strategy consists of sampling six regularly-spaced fields centred at Galactic longitudes of $\ell$\,=\,10$\degr$, 20$\degr$, 30$\degr$, 40$\degr$, 50$\degr$ and 60$\degr$, with each field covering approximately 5$\degr$ in longitude and 1$\fdg$7 in latitude. This observing strategy preserves many of the goals of a blind survey, producing a relatively unbiased data set, while also containing multiple significant features of Galactic structure such as the tangents of the Scutum--Centaurus and Sagittarius spiral arms and major star-forming regions such as W31, W43, and W51. By limiting the area coverage, JPS was also able to achieve significantly increased depth compared to existing sub-millimetre continuum surveys covering the same region (e.g. $\sim$\,$\times$10 deeper than ATLASGAL; Fig.~\ref{peakdist}).
Full details of the JPS observing strategy and preliminary results from the  region of $\ell$\,=\,27$\degr$\,--\,33$\degr$ can be found in \citet{Moore15}.

As part of the JLS, the JCMT also observed the outer Galactic Plane, encompassing Galactic longitudes in the range $\ell$\,=\,120$\degr$--240$\degr$ with the SCUBA-2 Ambitious Sky Survey (SASSy; Thompson et al., in preparation) with the range of $\ell$\,=\,60$\degr$--120$\degr$ covered by a separate project. These two projects, along with JPS, give complete coverage of the Galactic Plane visible from the JCMT at a constant mass sensitivity of roughly 100\,M$_{\sun}$ at 20\,kpc.

The structure of the paper is as follows: in Section 2, we present the observations and data reduction process of the JPS, with Section 3 describing the data. Section 4 describes the compact source catalogue, the extraction process, completeness tests and source properties. The data access is described in Section 5, with the content of the image data presented in Section 6. Comparisons with other Galactic Plane studies are made in Section 7, with preliminary conclusions made on the star-forming content in the JPS. Finally, we give a summary of the paper and conclusions in Section 8.

\section{Observations \& Data Reduction}

\subsection{Observing strategy}

Each individual survey field is sampled using a regular grid (see Fig.\ 3 of \citealt{Moore15}) of eleven circular tiles with diameter of one degree, observed using the $\emph{pong3600}$ mode of SCUBA-2 \citep{Bintley14}. Each individual $\emph{pong3600}$ takes 40-45\,minutes to observe and reaches a pixel-to-pixel rms of $\sim$\,92\,mJy\,beam$^{-1}$ in the assigned weather band, when reduced with 3-arcsec pixels. The data were taken between 2012 June and 2015 January, with each tile observed between seven and twelve times, depending on weather conditions, to obtain uniform noise across each field.

Although SCUBA-2 observes the 450-$\upmu$m and 850-$\upmu$m bands simultaneously, the assigned weather bands for the project had sky opacity values of $\uptau_{220}$\,$\simeq$\,0.08\,--\,0.16 at 220\,GHz, meaning that conditions were not reliably photometric at 450\,$\upmu$m, with only the brightest sources detected at low sensitivity and with unreliable fluxes.  These data are available from the JCMT archive, along with the basic JCMT Legacy Release 1 (JCMT-LR1) reductions (\citealt{Bell14}; Graves et al., in preparation), which are addressed in Section~\ref{access}.  In the next section, we describe the bespoke reduction of the 850-$\upmu$m data used to produce science-grade emission maps for the JPS project.

\subsection{Data reduction}

An outline of the general data-reduction process used by the JPS project can be found in \citet{Moore15}.  The data for this public release have been re-reduced with some key altered parameters, the details of which are highlighted below.

The reduction process makes use of the {\sc smurf} software package \citep{Jenness11}, which can be found in the \emph{Starlink} suite. The command used, {\sc smurf:makemap}, makes use of the Dynamic Iterative Map-Maker, outlined by \citet{Chapin13}. 

The data for this public release, JPSPR1, have been re-reduced with some key changes from the reduction presented in \citet{Moore15}. The individual observations of each tile are first reduced using the process in the initial JPS paper, but mapped onto 3-arcsec pixels, as opposed to the 4-arcsec grid used in the prior paper. These individual tiles are then coadded and masked for the emission, in a process similar to that of \citet{Mairs15}. This masking process sets to zero all data below a flux value of zero, ensuring negative noise is removed (see \citealt{Chapin13} and \citealt{Mairs15} for more details on external masking). The mask produced is then used to suppress the inflation of noise into spurious positive emission. Using an external mask enables some large-scale structures to be retained in the final map. The disadvantage of this is that real, low-level emission may be removed, especially since the data reduction process masks out structures on scales larger than 480 arcsec (the size of the SCUBA-2 footprint)

Whereas \citet{Moore15} reduced individual $\emph{pong3600}$ observations separately and then assembled them in a mosaic after removing the low signal-to-noise edges of each tile, in these complete data, all the observations in a field are reduced at the same time. This procedure improves the reliability of the reduction of individual scans. In the JPSPR1, all the data are coadded without clipping the noisy edges of the tiles, which consist of data taken outside the fully-sampled region of each $\emph{pong3600}$ observation, where the scan direction is changing. In JPSPR1, the edges are retained to increase the signal-to-noise ratio (SNR) in the overlapping regions. The number of repeated observations and the tile overlaps mean that this edge noise is largely suppressed in the central regions of each JPS field. However, some unreliable signal remains that has to be identified and weeded out (see below).

The parameters used for the customised {\sc smurf:makemap} configuration file can be found in Appendix~\ref{smurfappend}. The final maps are calibrated in units of Jy\,beam$^{-1}$ using the standard observatory-determined calibration factor of 537\,$\pm$\,26\,Jy\,beam$^{-1}$\,pW$^{-1}$ which is monitored regularly during the observations via measurements of planets such as Uranus \citep{Dempsey13}.

\section{JPSPR1 Data}
\label{sec:data}

\begin{figure*}
\begin{tabular}{l}
\includegraphics[width=0.99\linewidth]{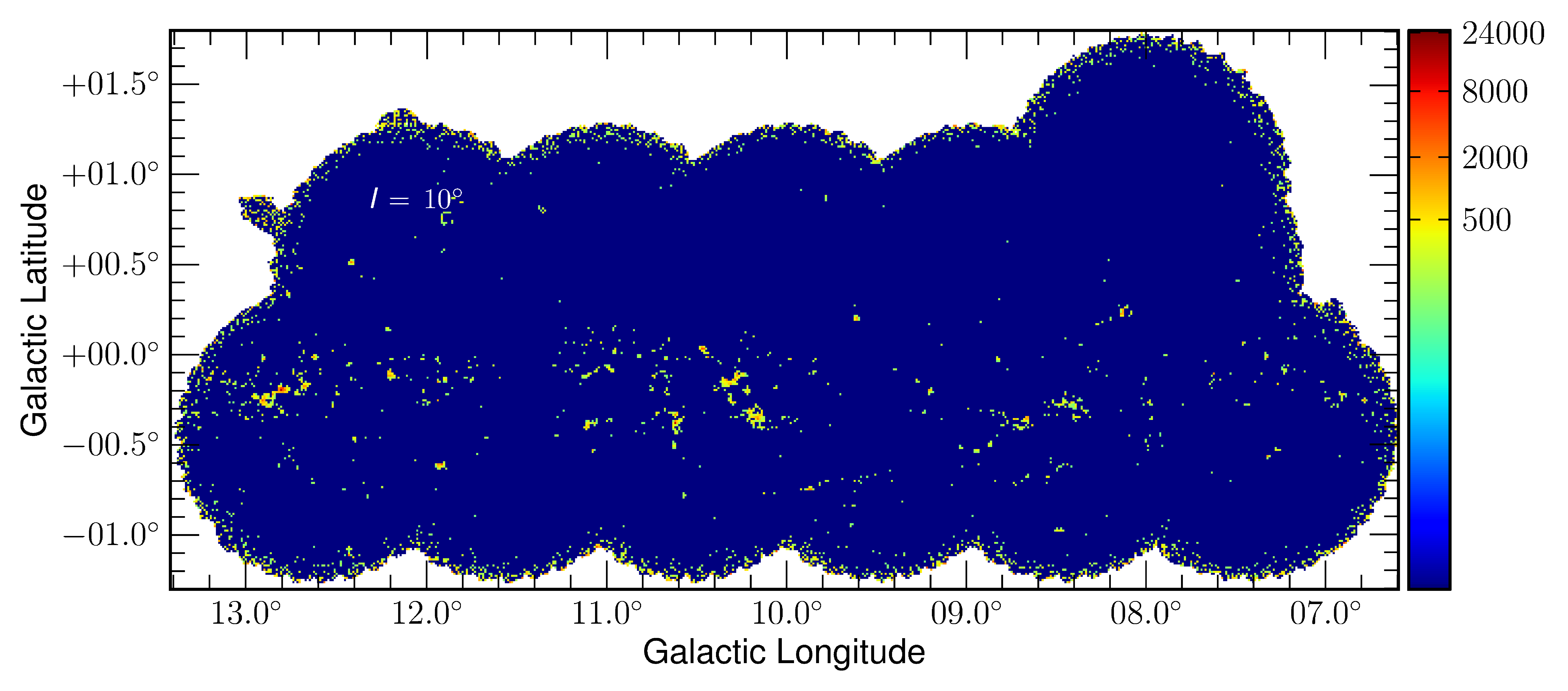}\\
\includegraphics[width=0.99\linewidth]{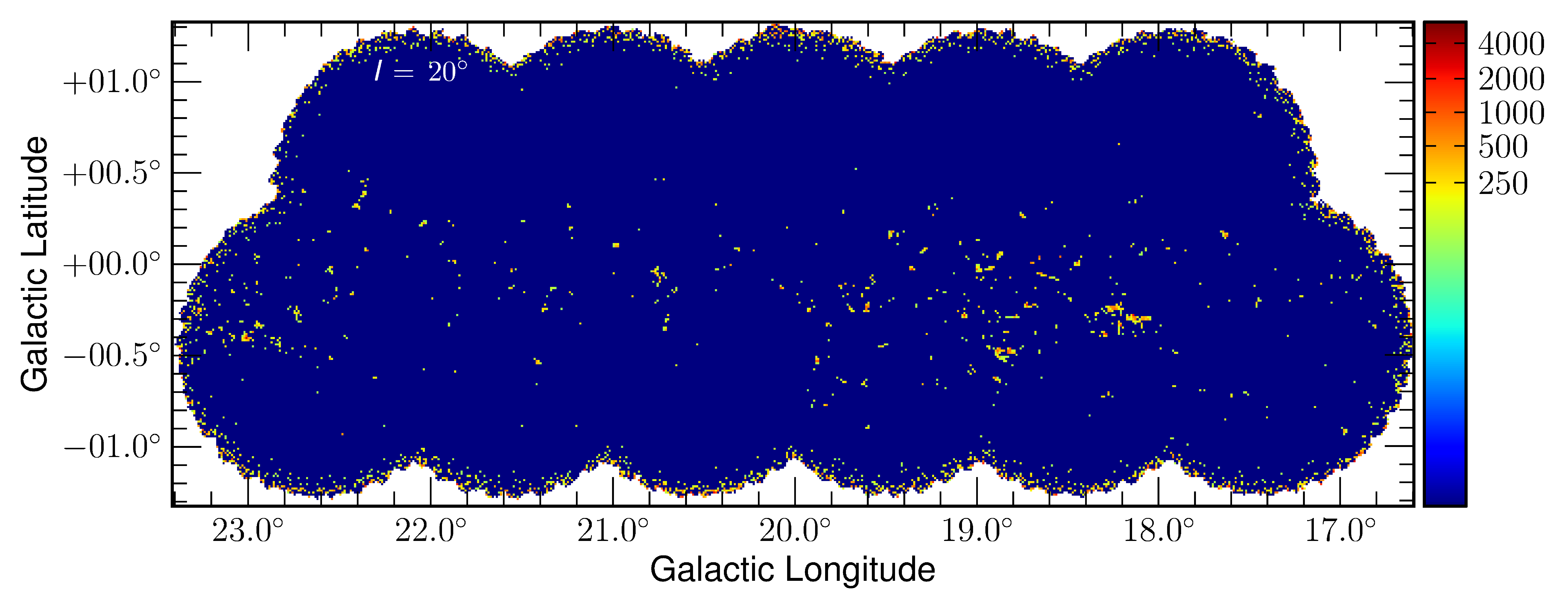}\\
\includegraphics[width=0.99\linewidth]{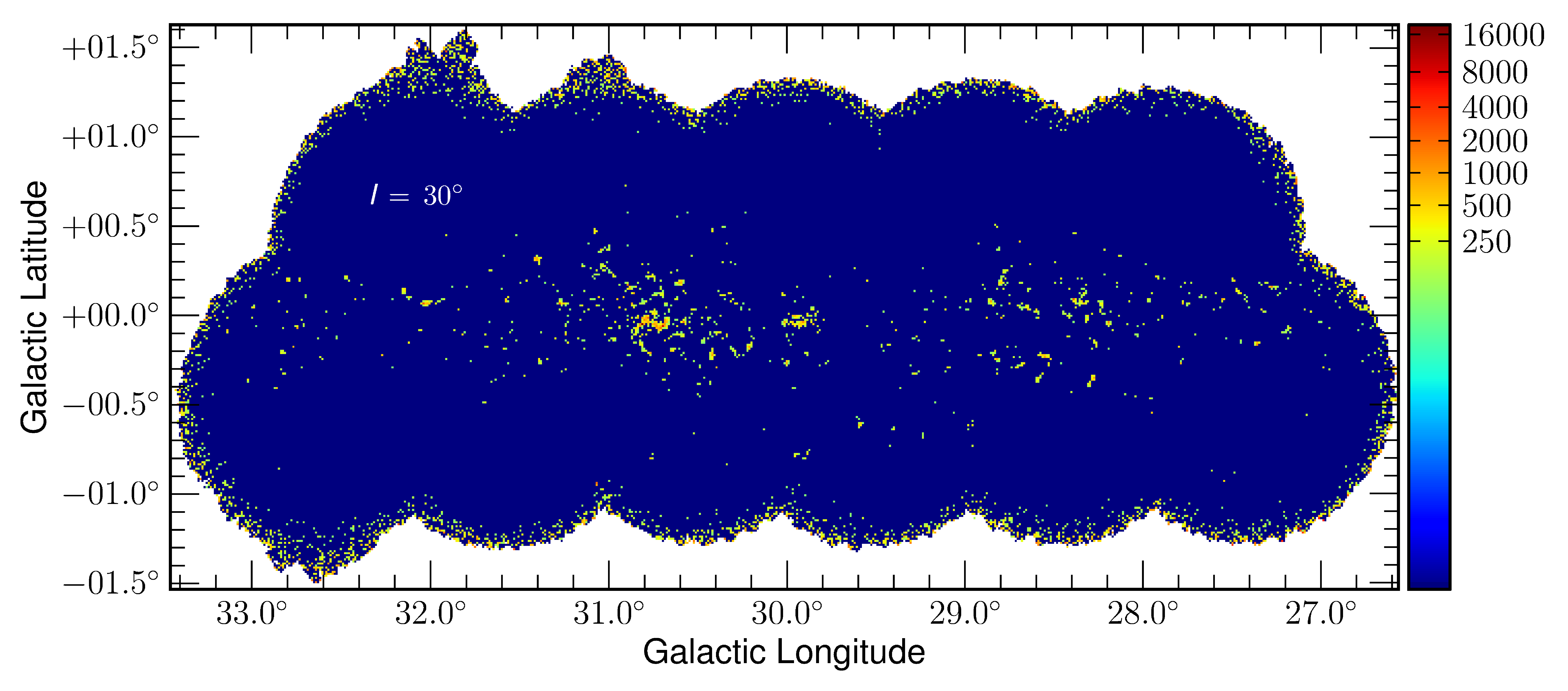}\\
\end{tabular}
\caption{The JPSPR1 maps for the first three fields, $\ell$\,=\,10$\degr$, 20$\degr$ and 30$\degr$. The intensity scale is in units of mJy\,beam$^{-1}$. Several areas can be seen where the SCUBA-2 instrument continued to take data beyond the edge of the standard circular $\emph{pong3600}$ tile.  These excursions are visible at the edges of most of the fields and the $\ell$\,=\,10$\degr$ field is misshapen in the top right tile. This extension is caused by the inclusion of a trial observation taken prior to the main survey that has a small positional offset from the standard grid pattern. Significant regions can be observed in each field with W31 found at $\ell$\,=\,18.25$\degr$,\,$\emph{b}$\,=\,-0.19$\degr$, W39 at $\ell$\,=\,18.86$\degr$,\,$\emph{b}$\,=\,-0.48$\degr$ and G29 and W43 found at $\ell$\,=\,29.95$\degr$,\,$\emph{b}$\,=\,-0.02$\degr$ and $\ell$\,=\,30.75$\degr$,\,$\emph{b}$\,=\,-0.05$\degr$, respectively.}
\label{dataimages}
\end{figure*}

\begin{figure*}
\begin{tabular}{l}
\includegraphics[width=0.99\linewidth]{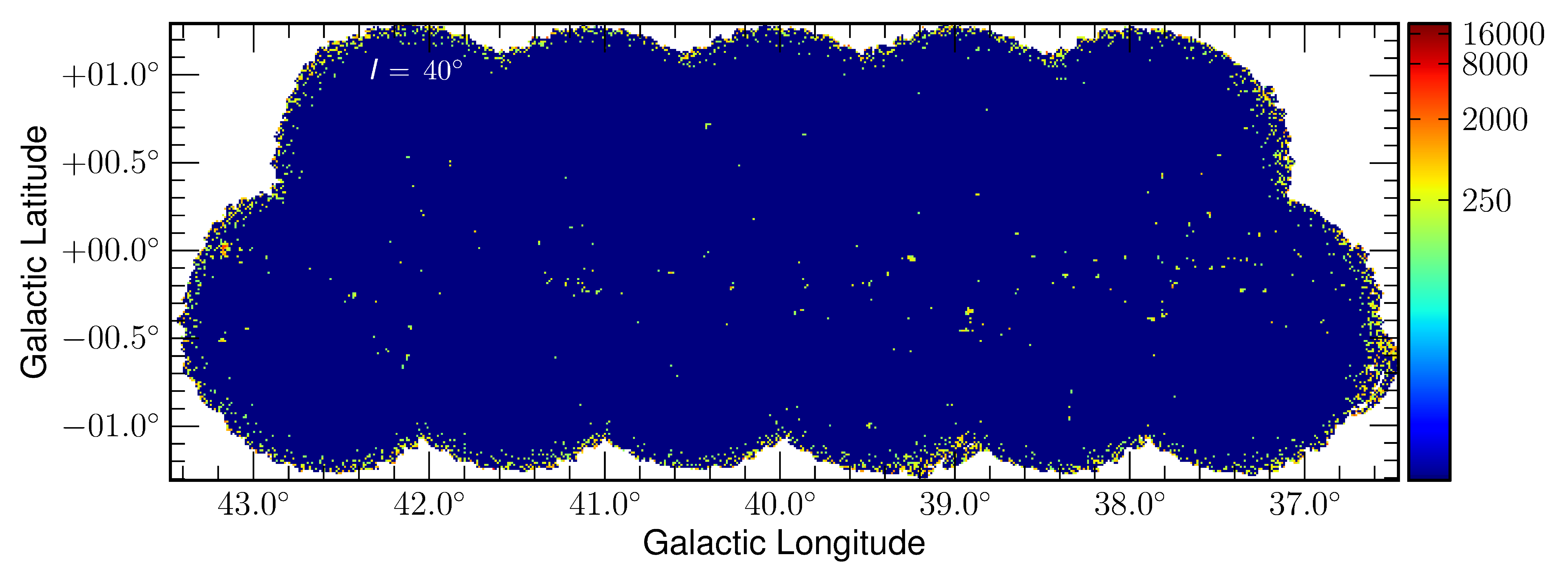}\\
\includegraphics[width=0.99\linewidth]{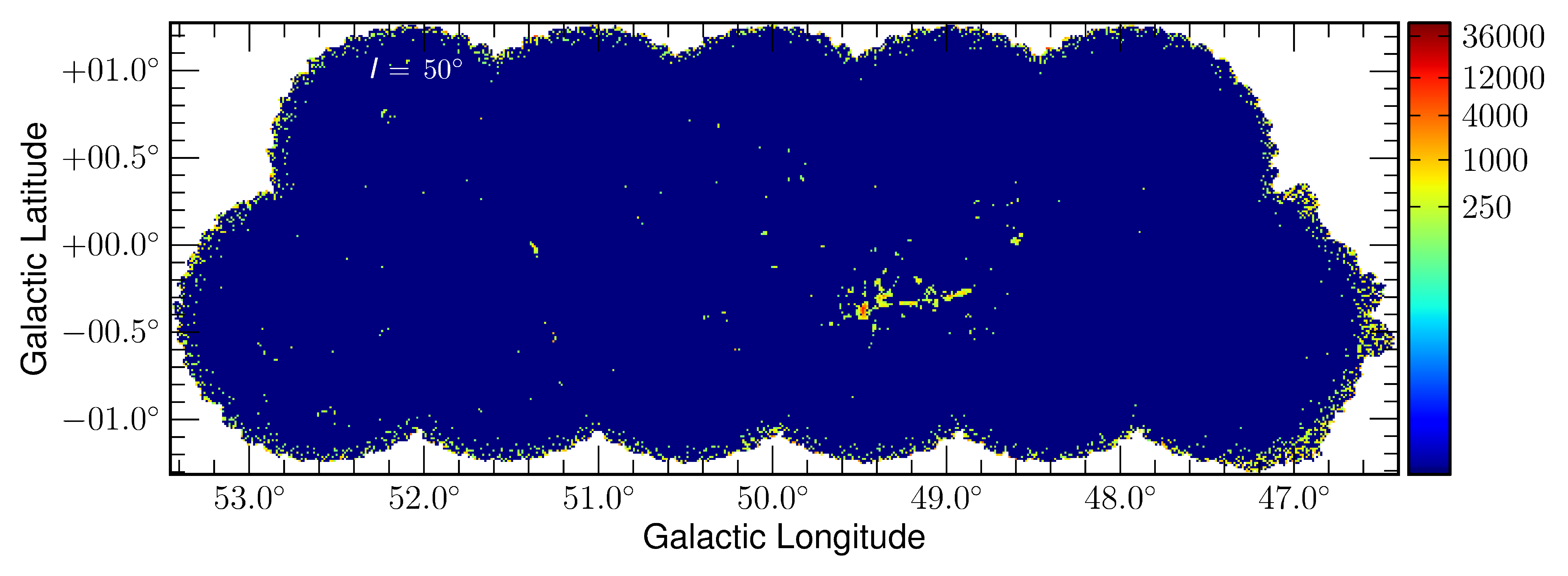}\\
\includegraphics[width=0.99\linewidth]{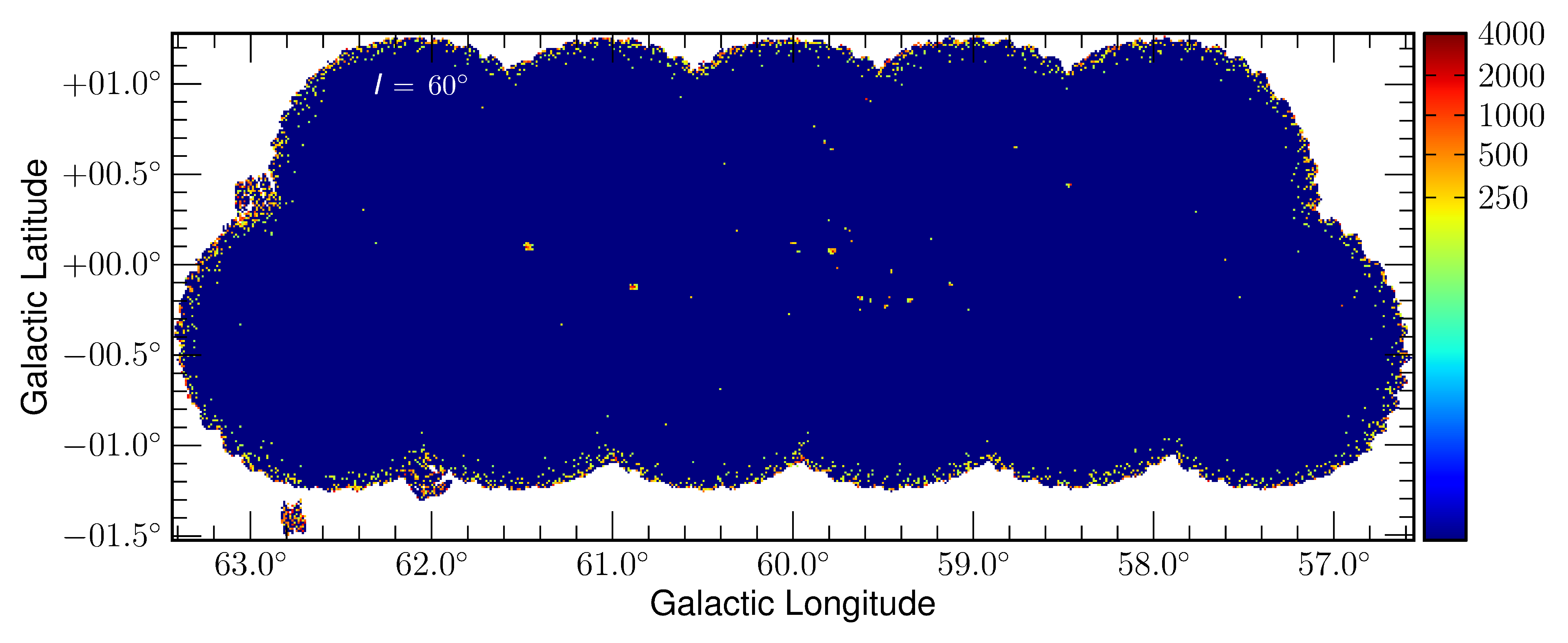}\\
\end{tabular}
\caption{The JPSPR1 maps for the three fields centred at $\ell$\,=\,40$\degr$, 50$\degr$ and 60$\degr$. The intensity scale is in units of mJy\,beam$^{-1}$. The W51 star-forming region can be located at $\ell$\,=\,49.40$\degr$,\,$\emph{b}$\,=\,-0.38$\degr$.}
\label{dataimages2}
\end{figure*}

Fig.~\ref{dataimages} shows the JPSPR1 image data for the first three JPS fields, with Fig.~\ref{dataimages2} showing the results in the $\ell$\,=\,40$\degr$, 50$\degr$ and 60$\degr$ fields. These images show only the brightest sources due to the dynamic range and image resolution. There are significant real compact and filamentary sources at lower surface brightness that will be discussed later.

The $\ell$\,=30$\degr$ and 50$\degr$ fields contain small regions of negative surface brightness (``negative bowling'') around the brightest sources, especially W43 and W51. Many of the details of the reduction process, in particular the selection of pipeline parameters and the masking process, aim to mitigate this effect but the presence of regions of negative bowling means that the source catalogues will not be complete in those regions.  However, while these sources are of significant interest, the problem affects a very small area (radius $\sim$\,100 arcsec around the sources) compared to the entirety of the JPS coverage. The maximum level of negative bowling is 6\,$\upsigma$ and 13\,$\upsigma$ around W43 and W51, respectively. This compares to the maximum data values of $\sim$ 700\,$\upsigma$ and $\sim$ 2100\,$\upsigma$ in the two regions, respectively. Effective correction of such artifacts will be the subject of a future study.

Histograms of flux values per pixel in each of the six fields, along with Gaussian fits to the respective distributions, are displayed in Fig.~\ref{pixhisto} in Appendix~\ref{pixhistoappend}. In each of these histograms, there is a slight negative excess with respect to the normal distribution, caused by non-Gaussian noise in the wings, while the positive excess comes primarily from the real astronomical signal.  The rms noise for each field is estimated from the latter fits and the resulting values are displayed in Table~\ref{rmsvalues}.

An alternative measure of the noise is contained in the variance data produced by the data-processing and reduction software, based on the signal variation in each multiply-sampled spatial pixel.  The resulting variance maps are shown in Appendix~\ref{varianceappend}.

Since the square root of the variance is equivalent to the standard deviation of the noise, we can compare the resulting alternative standard deviation values to those obtained from pixel-to-pixel variations in the reduced data found above. Histograms of these standard deviation values for the pixels in each field are shown in Fig.~\ref{noisehisto}.  The pixel-to-pixel rms values calculated from Fig.~\ref{pixhisto} are overlaid and each distribution peaks close to the latter. The variance-derived standard deviation histograms have a similar profile to those of the data.

\begin{table}
\begin{center}
\caption{The values for rms for each field determined from the Gaussian fits in Fig.~\ref{pixhisto}, with the comparison to the target rms after smoothing over the beam. The $\upsigma_{\rmn{rms}}$ numbers are pixel-to-pixel noise values which are dependent on pixel size, whereas the smoothed numbers are beam-to-beam. The pixel-to-pixel values are those used from this point forward.}
\label{rmsvalues}
\begin{tabular}{lcccc}\hline
Field & $\upsigma_{\rmn{rms}}$ & Smoothed rms & Source & Sources\\
 & (mJy\,beam$^{-1}$) & (mJy\,beam$^{-1}$) & Numbers & per deg$^{-2}$\\
\hline
$\ell$ = 10$\degr$ & 31.06 & 8.42 & 1,883 & 181\\
$\ell$ = 20$\degr$ & 28.67 & 7.76 & 1,773 & 170\\
$\ell$ = 30$\degr$ & 29.89 & 8.17 & 2,149 & 207\\
$\ell$ = 40$\degr$ & 27.89 & 7.15 & 925 & 89\\
$\ell$ = 50$\degr$ & 25.66 & 5.98 & 822 & 79\\
$\ell$ = 60$\degr$ & 26.40 & 5.66 & 261 & 25\\
\hline
\end{tabular}
\end{center}
\end{table}

\section{Compact Source Catalogue}

\subsection{Compact source detection}

Extraction of the compact sources from the JPSPR1 data was done using the {\sc FellWalker} (FW; \citealt{Berry15}) source-extraction algorithm, part of the $\emph{Starlink}$ {\sc cupid} package \citep{Berry07}. Details of this choice of source-extraction process can be found in \citet{Moore15}.

FW is most effective at extracting clumps when the background noise is distributed uniformly. As a result, we have used the same method as employed in \citet{Moore15} and \citet{Rigby16}, by performing the source extraction on a signal-to-noise ratio (SNR) map. This is produced using the intensity data and the variance maps shown in Appendix~\ref{varianceappend}. 

All sources containing emission above a threshold of 3\,$\upsigma$ (i.e., three times the pixel-to-pixel noise) in the SNR map are initially identified using FW as part of the task {\sc cupid:findclumps}. This creates a mask that is applied to the intensity map as input for the task {\sc cupid:extractclumps}, which extracts the peak and integrated flux density values that are reported in the source catalogue. A further threshold for {\sc cupid:findclumps} was the minimum number of contiguous pixels to qualify as a genuine source.  This was set at 12, which is the number of pixels expected to be found in an unresolved source with a peak SNR of 5\,$\upsigma$, given a 14.5-arcsec beam and 3-arcsec pixels. The other parameters used in the FW process are given in Appendix~\ref{fellparamsappend}.

Further thresholds to membership of the final catalogue are a lower limit to the peak SNR of 5\,$\upsigma$ and an aspect-ratio cut, ensuring only reliable compact sources are included. The latter is necessary because the fidelity of the JPS data to extended (i.e., filamentary) structures has not yet been quantified and because extraction of extended sources is a complex problem, the approach to which is likely to depend strongly on the intended science. Therefore, sources with aspect ratios (the ratio of major to minor axis size) greater than 5.0 are rejected. 

A further cut was required to account for sources found by {\sc fellwalker} near the edges of the field that are the result of noise or are undersampled and so have unreliable fluxes. These sources were cut by extracting the values from the variance image at the positions of the catalogued sources. As can be seen from the variance images in Fig.~\ref{varianceimages}, the edges of these maps have $\sim$ 8 times the value of the areas in the centres of the maps. Therefore, sources with high peak values in these variance-extracted clumps are likely to be found on the edge or to be otherwise unreliable, i.e. a cut in the value of the variance at the source position at $\sim$\,3000\,mJy\,beam$^{-1}$. Application of this criterion resulted in a total of 7,813 sources found away from the edges of the maps.

Table~\ref{sourcecatalogue} contains a small part of the 7,813 sources found in the final JPSPR1 compact source catalogue, with the numbers found in each region listed in Table~\ref{rmsvalues}, along with the sources per square degree. The full source catalogue can be accessed from the CANFAR archive as well as listed in the Supporting Information.

The catalogue is made up of mainly star-forming regions, those which are both starless and protostellar. There will, however, be sources of a different nature included in the catalogue. We estimate that $\sim$ 3.5 per cent of sources are likely to be evolved stars, after determining the number of known AGB stars found in the JPS longitude range \citep{Suh11}. We do not, however, expect to detect any sources of a cosmological nature \citep{Geach17} due to the JPS sensitivity. The nature of sources in the catalogue will be determined in a future study when the individual sources are compared to the molecular environment and star-formation tracers but sources already detected by ATLASGAL will be known.

\subsection{Recovered flux densities}

\begin{figure}
\begin{tabular}{l}
\includegraphics[scale=0.50]{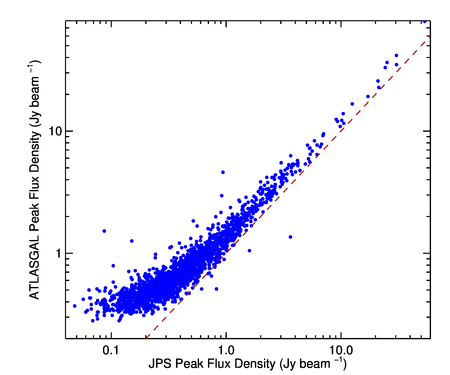} \\
\includegraphics[scale=0.50]{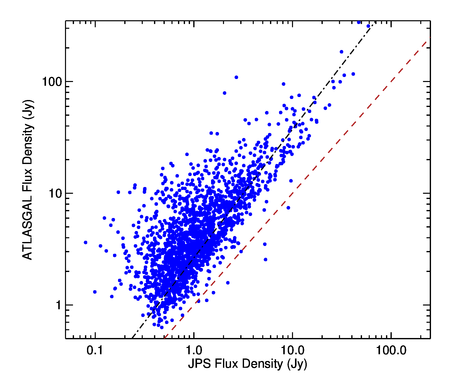} \\
\end{tabular}
\caption{Comparison of the JPS sources with the spatially corresponding ATLASGAL clumps. Top panel: peak flux densities. Bottom panel: integrated flux densities. The red dashed lines correspond to the 1:1 line, with the black dot-dash line representing the linear fit to the matched integrated flux densities.}
\label{ATLASmatches}
\end{figure}

As a check on the flux densities of the sources recovered by {\sc fellwalker}, we positionally matched the JPS sources with ATLASGAL compact sources \citep{Contreras13,Urquhart14}. The peak positions in the ATLASGAL catalogue were used to search for the nearest JPS sources within a radius of 19 arcsecs, equivalent to the APEX 870-$\upmu$m beam, finding 1,918 matches. The comparison of the JPS and ATLASGAL peak flux densities are shown in the top panel of Fig.~\ref{ATLASmatches}.  The results show a clear linear relationship with a small systematic offset to lower JPS peak flux densities.  This offset is expected and can be accounted for by the smaller JPS beam and the fact that most sources are somewhat resolved.  This effect is explored in \citet{Moore15}, who also demonstrate that the appropriate correction can be obtained by smoothing the JPS data to the ATLASGAL resolution.  

\begin{figure}
\includegraphics[scale=0.50]{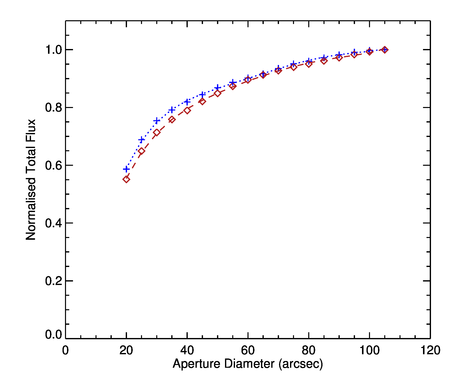}
\caption{Aperture corrections determined for Neptune (blue crosses) and a co-add of three Uranus observations (red diamonds), with the values normalised to the maximum value. A fit to each distribution (Neptune: blue, dotted line; Uranus: red dashed line) with a fifth-order polynomial, from which an average of the two was applied to the integrated flux densities for the JPS compact sources.}
\label{aperture}
\end{figure}

\begin{figure*}
\begin{tabular}{ll}
\includegraphics[width=0.5\linewidth]{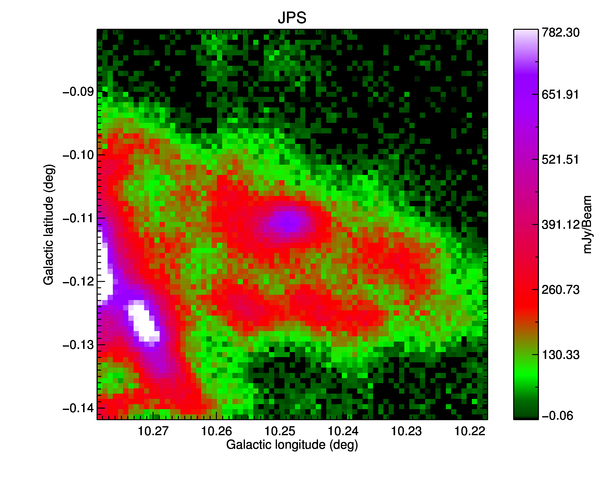} & \includegraphics[width=0.5\linewidth]{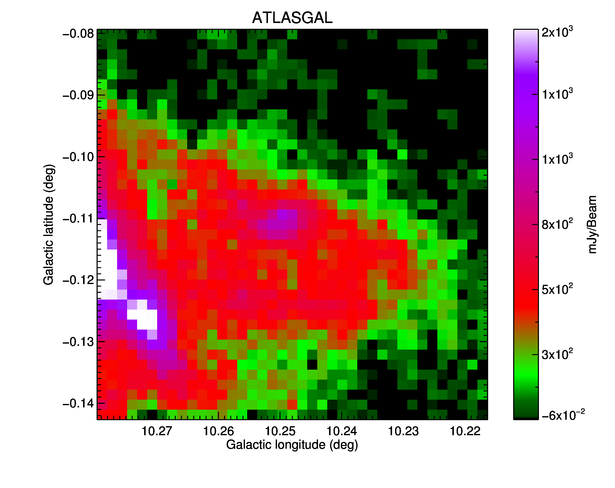} \\
\end{tabular}
\caption{Comparison images of positionally matched JPS and ATLASGAL sources. The left panel is the JPS source, the right panel is the ATLASGAL source. The source is centred at $\ell$ = 10.248, $\emph{b}$ = -0.111.}
\label{ATLAS_JPS_comp_images}
\end{figure*}

\begin{figure}
\includegraphics[scale=0.50]{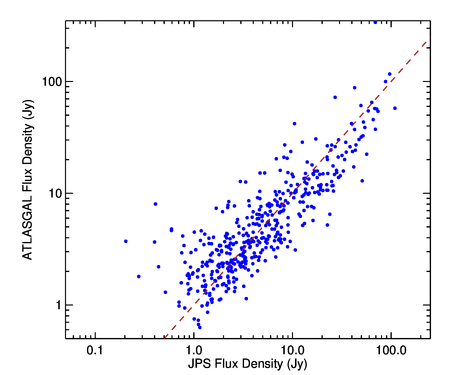}
\caption{The comparison of the peak and integrated flux densities for sources extracted in the smoothed JPS $\ell$ = 10$\degr$. The red dashed line indicates the 1:1 line.}
\label{ATLAS_smooth}
\end{figure}

As the FW algorithm essentially uses aperture photometry, with the aperture size and shape set by the number and distribution of contiguous pixels above the desired detection threshold, the integrated flux densities will be affected by the loss of signal outside this aperture and below the threshold. As shown by \citet{Dempsey13}, the wings of the JCMT beam contain significant power. Since the effective aperture is SNR-dependent, this loss can be therefore quite significant for the fainter sources. 

To measure the JCMT beam shape and obtain an aperture correction, one observation of Neptune and three of Uranus were made during the survey and reduced in the same manner as the JPS data.  The percentage of the total flux found in increasing circular apertures is shown for both planets in Fig.~\ref{aperture}. Both sets of points are fitted with an interpolating fifth-order polynomial producing two aperture-correction curves, and an average of the two is applied to the integrated flux densities produced by FW. The correction, $C_{\rmn{int}}$ is given by:

\begin{equation}
\begin{split}
C_{\rmn{int}} & = -0.329 + 0.077R_{\rmn{eff}} + 2.169\,\times10^{-3}\,R_{\rmn{eff}}^{2} + 3.177\,\times10^{-5}\,R_{\rmn{eff}}^{3} \\
 & - 2.301\,\times10^{-7}\,R_{\rmn{eff}}^{4} + 6.507\,\times10^{-10}\,R_{\rmn{eff}}^{5}
\end{split}
\end{equation}

\noindent where $R_{\rmn{eff}}$ is the effective radius in units of arsec. The application of a circular aperture correction to non-circular sources is, of course, approximate but the sources most affected are the fainter ones, which tend to be more compact, as only the peaks are detected. Larger sources are less affected, being generally extended in at least one direction and the correction is negligible for brighter sources of any shape. This correction is consistent with that of \citet{Dempsey13}.

Further evidence for the power in the wings of the JCMT beam is the ratio of integrated-to-peak flux density of Neptune, which was calculated to be 28\,$\pm$\,1 (a Gaussian would have a ratio of $\sim$ 15), with a FWHM size of 4.8 pixels, corresponding to 14.4 arcsec. As Neptune can be considered as a true point source at a size of $\sim$1 arcsec, this size of 14.4\,$\pm$\,0.3 arcsec is taken to be the half-power width of the beam, which is consistent with the assumed beam size. From this measurement, the beam integral or oversampling factor is taken to be 28 pixels per beam.

The bottom panel of Fig.~\ref{ATLASmatches} compares the integrated flux densities of the 1,918 positionally matched JPS and ATLASGAL sources. There is a clear correlation but, with the exception of a few sources that lie close to the 1:1 line, the JPS flux densities are consistently lower than those of the corresponding ATLASGAL sources by a larger factor than that affecting the peak flux densities, with a mean ratio of 0.298\,$\pm$\,0.004 and a median of 0.283.  The linear best fit to the relationship has a gradient of 1.15\,$\pm$\,0.16, the large uncertainty resulting from the considerable scatter in the flux-density ratios. 

Visual inspection of sources in both catalogues, an example of which is shown in Fig.~\ref{ATLAS_JPS_comp_images}, reveals that the ATLASGAL surface-brightness distribution tends to be broken up into multiple components by the higher angular resolution of the JPS data, and substructures are identified as separate sources in the latter. Since the source matching is one-to-one, the JPS components mostly have lower integrated flux densities than the ATLASGAL sources. As seen in Fig.~\ref{peakint}, there are very few real point sources, so it is expected that only the few faint isolated point sources would appear on the 1:1 line.

A check on these recovered integrated flux densities was made by smoothing the $\ell$ = 10$\degr$ JPS field to the resolution of the ATLASGAL data, 19 arcsecs. Replicating the JPS source extraction on this smoothed field and then positionally matching to the ATLASGAL data found 434 matches.  The comparison of integrated flux densities is shown in Fig.~\ref{ATLAS_smooth}. The sources now lie along the 1:1 line with reduced scatter, a mean ratio of 1.110\,$\pm$\,0.002 and median of 1.033, indicating that the improved resolution and sensitivity of the JPS data reveals substructure in the ATLASGAL sources, breaking most of them up into multiple compact sources.  Any residual discrepancy is likely to be the result of the different source-extraction methods and detection thresholds used in the two surveys.

As discussed in \citet{Moore15}, the 850-$\micron$ data can be subject to contamination from CO and free-free emission. $^{12}$CO $J = 3-2$ contamination, however, was found to be generally at the level of a few per cent in the $\ell$\,=\,30$\degr$ field, consistent with other results \citep[e.g.][]{Wyrowski06,Schuller09,Drabek12}, becoming more significant in sources with strong, optically thin outflow wings. For example, the JCMT Gould Belt Survey found an average of $\sim$ 17 per cent towards sources in Orion A \citep{Coude16}. Free-free contamination from ionised gas would mostly affect the brightest sources containing H{\sc ii} regions, such as W43 and W51, but previous studies have found that this contributes less than $\sim$ 20 per cent in W43 \citep{Schuller09} and 12 per cent in W40 \citep{Rumble16}.

\subsection{Completeness tests}
\label{complete}

The completeness of each JPS field as a function of peak source flux density was estimated by repeatedly injecting compact artificial sources into the JPS fields. The source extraction process was repeated on these new artificial+real fields and the recovered source numbers were compared to both the real and artificial catalogues. To minimise non-linear effects caused by artificial sources blending with each other as well as with real sources, each injection of fake sources was limited to 10 per cent of the number of real sources found in that particular JPS field. This 10 per cent injection was repeated until 10,000 artificial sources in total had been injected into each JPS field.

The artificial sources were produced and injected using the {\sc cupid:makeclumps} routine. The sources have Gaussian profiles with FWHM of 7 pixels (21 arcsec) in both $\ell$ and $\emph{b}$, equal to the peak of the source size distribution including rejected sources (see Fig.~\ref{sourcesize}), and were distributed uniformly across $\ell$ and $\emph{b}$, and had a uniform flux distribution between 2 and 500\,mJy\,beam$^{-1}$.

\begin{figure}
\includegraphics[scale=0.50]{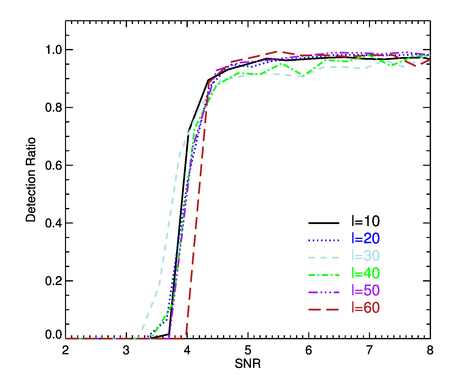}
\caption{The recovery fraction of injected artificial sources in each JPS field as a function of signal-to-noise ratio.}
\label{recovratio}
\end{figure}

The recovery fraction in each JPS field as a function of SNR ratio is shown in Fig.~\ref{recovratio}. The fraction is approximately 95 per cent or above for sources with peak flux densities greater than 5\,$\upsigma$ in four of the six JPS fields, whereas the rate in the $\ell$\,=30$\degr$ and 40$\degr$ fields is approximately 90 per cent at 5\,$\upsigma$. The latter two fields reach 95 per cent completeness at approximately 7.5\,$\upsigma$ and 6.5\,$\upsigma$, respectively.

\begin{figure}
\includegraphics[scale=0.50]{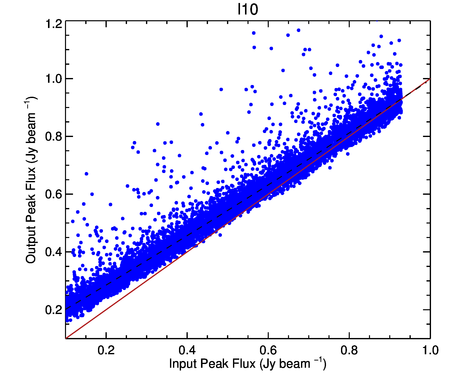}
\caption{Comparison of injected and recovered peak flux densities in the $\ell$\,=10$\degr$ region. The solid red line represents the line of equality and the black dashed line is a least-squares fit to the trend.}
\label{recovcomppeak}
\end{figure}

The comparison of injected and recovered peak flux densities in the $\ell$\,=10$\degr$ region is displayed in Fig.~\ref{recovcomppeak}. Overlaid on the plot are a least-squares fit to the data and the 1:1 line. The injected and recovered flux densities are generally well correlated but below $\sim$ 0.9\,Jy the recovered flux densities are systematically higher than the those of the injected sources. As the injected sources are added to the real data, they fall on top of real noise. As the FW routine assigns the peak source position to the pixel with the highest signal, any positive noise added onto the source flux creates a positive bias. This bias boosts the recovered flux density by approximately 1\,$\upsigma$ but becomes less of an effect after $\sim$ 0.9\,Jy. This trend towards artificially boosted recovered flux densites is well fitted by a second-order polynomial of the form:

\begin{equation}
S_{\rmn{peak}}  = a + bS_{\rmn{u}} + cS_{\rmn{u}}^{2}
\end{equation}

\noindent where $a$, $b$, and $c$ are listed in Table~\ref{peakfits} and $S_{\rmn{u}}$ is the uncorrected peak flux density.

\begin{table}
\begin{center}
\caption{The parameters of the fits and the limit to the corrections, calculated as the point where the second-order polynomial crosses the 1:1 line.}
\label{peakfits}
\begin{tabular}{lcccc}\hline
Field & $a$ & $b$ & $c$ & Limit (Jy)\\
\hline
$\ell$ = 10$\degr$ & 0.119 & 0.822 & 0.056 & 0.948 \\
$\ell$ = 20$\degr$ & 0.112 & 0.817 & 0.061 & 0.848 \\
$\ell$ = 30$\degr$ & 0.117 & 0.810 & 0.065 & 0.878 \\
$\ell$ = 40$\degr$ & 0.113 & 0.821 & 0.060 & 0.917 \\
$\ell$ = 50$\degr$ & 0.102 & 0.830 & 0.051 & 0.783 \\
$\ell$ = 60$\degr$ & 0.107 & 0.814 & 0.066 & 0.804 \\
\hline
\end{tabular}
\end{center}
\end{table}

Since this ``flux boosting'' affects real sources as well as artificial ones, the fit is used to correct the fluxes of all sources with a peak flux density below the cutoff listed in Table~\ref{peakfits}.  Sources above this cutoff are not required to be corrected as the correction is small, and less than the flux calibration uncertainties in 850-$\upmu$m SCUBA-2 data ($\sim$5 per cent, \citealp{Dempsey13}). The scatter of sources with extracted flux densities much greater than the trend in Fig.~\ref{recovcomppeak} occurs where the injected sources fall on top of an existing real source, affecting less than two per cent of injected sources.

Tests for false-positive detections were done by multiplying the images by $-1$ and running the FW procedure on the inverted data, using the same FW parameters. The only false-positive sources found in this way were located in regions of negative bowling observed in the $\ell$\,=30$\degr$ and 50$\degr$ fields, as discussed in Section~\ref{sec:data}. The lack of false positives, despite the number of pixels with negative values (Fig.~\ref{pixhisto}), is due to the minimum pixel criterion in the FW parameters. Namely, ``spikes'' with high SNR in the inverted maps are too isolated to be detected by FW, other than in regions of significant negative bowling, such as those found around W43 and W51.

\subsection{Angular size distribution}

\begin{figure}
\begin{tabular}{l}
\includegraphics[scale=0.50]{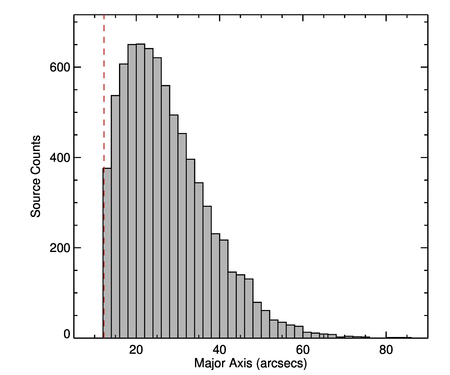} \\
\includegraphics[scale=0.50]{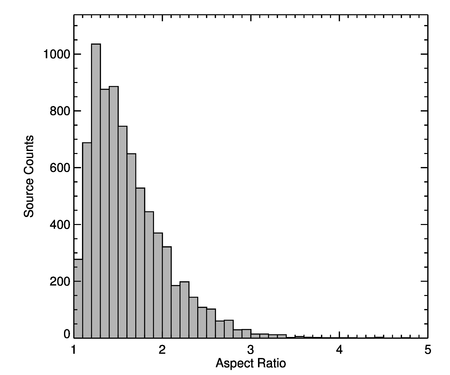} \\
\end{tabular}
\caption{Top panel: The distribution of major-axis sizes for the sources catalogued in the JPS. The red, dashed line indicates the 1-$\upsigma$ width of the 14.4\,arcsec-JCMT beam. Bottom panel: The distribution of aspect ratios of the catalogued sources.}
\label{sourcesize}
\end{figure}

The source size distribution, indicated by the source full major axis, is displayed in the top panel of Fig~\ref{sourcesize}, together with the 1-$\upsigma$ width of the JCMT beam size at 850\,$\upmu$m, with the beam size found to be 14.4\,arcsec (see above). The source sizes plotted are the 1-$\upsigma$ width as this is the parameter measured by FW.

The lower panel of Fig~\ref{sourcesize} contains the distribution of aspect ratios of the JPS sources, which has a median value of 1.51, consistent with that found for ATLASGAL \citep{Contreras13} and the BGPS \citep{Rosolowsky10}.

\begin{figure}
\begin{tabular}{l}
\includegraphics[scale=0.50]{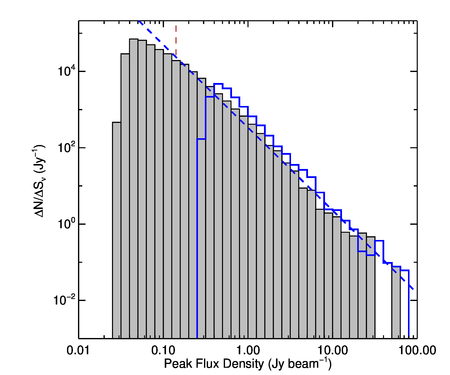} \\
\includegraphics[scale=0.50]{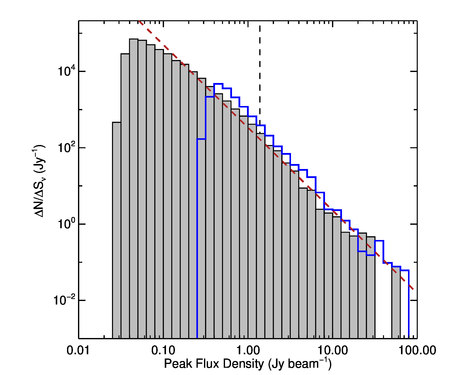} \\
\end{tabular}
\caption{Peak and integrated flux density distributions for the JPS (grey filled histogram) compared to the ATLASGAL distribution (blue histogram) in the top and bottom panel, respectively. The dashed red line indicates the least-squares fit to the JPS distribution.}
\label{peakdist}
\end{figure}

\begin{figure}
\includegraphics[scale=0.50]{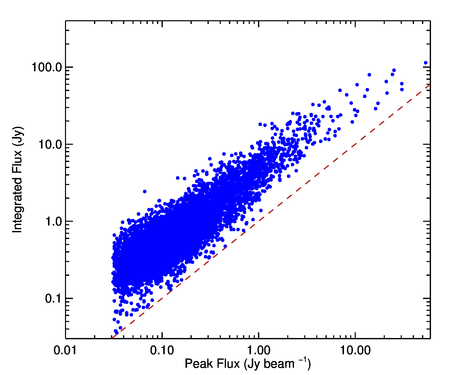}
\caption{Comparison of the peak and integrated flux densities of compact sources extracted from the JPS data. The red dashed line indicates the 1:1 relation, the locus at which true point sources would be found.}
\label{peakint}
\end{figure}

\subsection{Peak \& integrated flux density distributions}

By normalising the peak flux of each source to a multiple of \,$\upsigma_{\rmn{rms}}$, we can assume that the peak flux distributions in each of the 6 regions are drawn from the same population via a Kolmogorov--Smirnov (KS) test at the 1\,$\upsigma$ level. As a result, the catalogues for the individual regions can be combined, with the peak and integrated flux distributions shown in Fig.~\ref{peakdist}.

Assuming the flux-density distributions to be single power laws above the turnovers, represented by the expression $\Delta$$\emph{N/}$$\Delta$$S_{\nu}$\,$\propto$\,$S^{-\alpha}$, they are fitted by linear least-squares with values for $\alpha$\,=\,2.24\,$\pm$\,0.12 and 2.56\,$\pm$\,0.18 for the peak and integrated distributions, respectively. These values are consistent with those found for ATLASGAL (2.43\,$\pm$\,0.04 and 2.30\,$\pm$\,0.06, respectively; \citealt{Contreras13}).

These distributions are also compared to those of the ATLASGAL catalogue over the same area as the JPS, indicated by the blue histogram \citep{Contreras13,Urquhart14}. The comparison of the two surveys shows similar distributions with consistent power-law slopes for the peak  and integrated flux densities but a turnover indicating the completeness limit, and hence the sensitivity of the survey, being a factor of $\sim$ 10 deeper in JPS than ATLASGAL.

The comparison between peak and integrated flux density for individual JPS compact sources is shown in Fig.~\ref{peakint}, with the 1:1 line for reference. The range of integrated-to-peak flux density ratios is $\sim$ 1--37; however, sources are generally found within $S_{\rmn{int}}/S_{\rmn{peak}}$\,$\sim$\,3--5, indicating that very few true point sources are found in the JPS data.

\section{Data Access \& Products}
\label{access}

The JPS data are available to download from the CANFAR archive\footnote{http://www.canfar.phys.uvic.ca/vosui/\#/JPSPR1}. The data are presented in the FITS format and are available as mosaicked maps of the six separate JPS fields. In addition to these maps, the variance noise maps are also presented to allow the user to create their own SNR maps, with external masks also provided.
The raw data can be downloaded from the Canadian Astronomy Data Centre's JCMT Science Archive\footnote{http://www.cadc-ccda.hia-iha.nrc-cnrc.gc.ca/en/jcmt/} using the Project ID MJLSJ02.

As part of the JCMT Legacy Release 1 (JCMT-LR1), these data have also been processed, along with all other SCUBA-2 850\,$\upmu$m JCMT Legacy Survey data using the observatory's own reduction configuration (\citealt{Bell14}; Graves et al., in preparation). The JCMT-LR1, however, used a generic data reduction and source extraction procedure which is greatly improved upon in this JPS data release due to the customised process outlined above. In parallel, the JCMT Gould Belt Survey found that their customised reduction, which was similar to that of the JPS in using the external masking method, resulted in the detection of significantly more extended emission \citep{Mairs15}, 

\section{Example JPS Data}

\begin{landscape}
\begin{figure}
\begin{center}
\begin{tabular}{llll}
\includegraphics[width=0.33\textwidth]{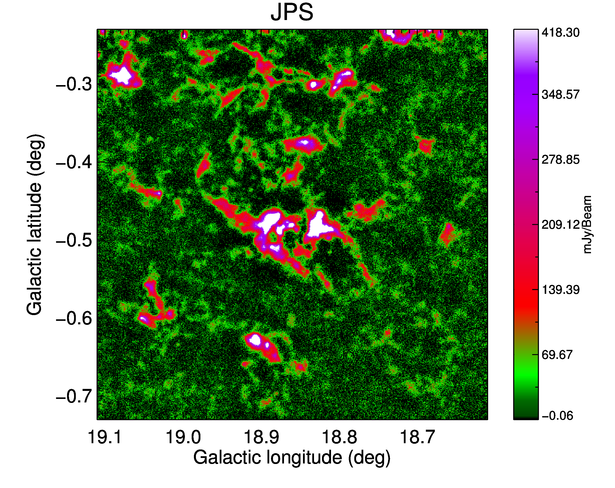} & \includegraphics[width=0.33\textwidth]{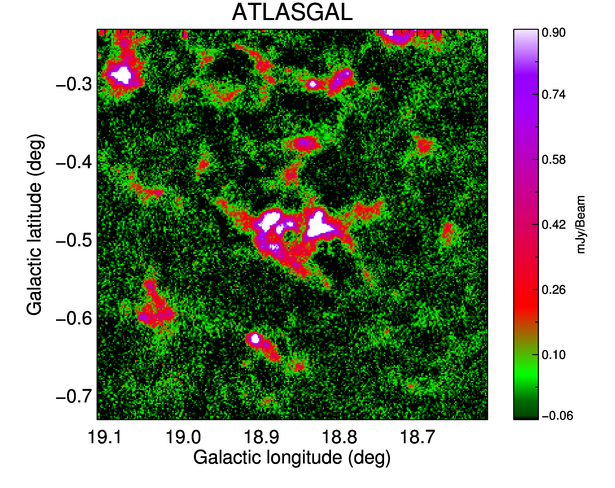} & \includegraphics[width=0.33\textwidth]{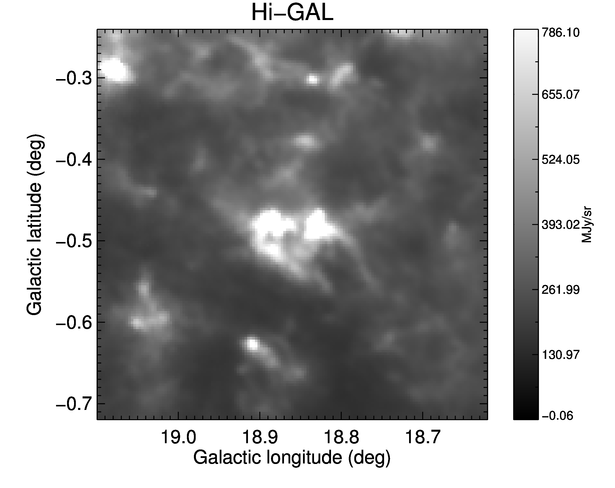} & \includegraphics[width=0.33\textwidth]{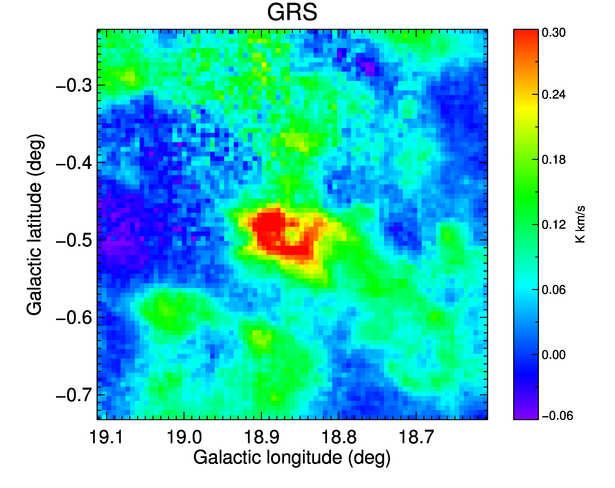} \\
\includegraphics[width=0.33\textwidth]{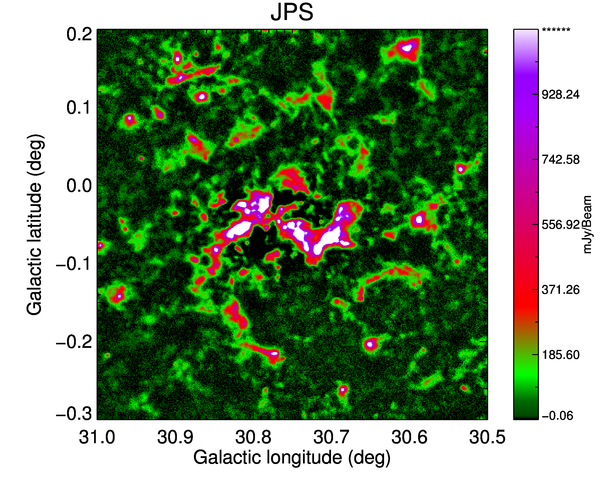} & \includegraphics[width=0.33\textwidth]{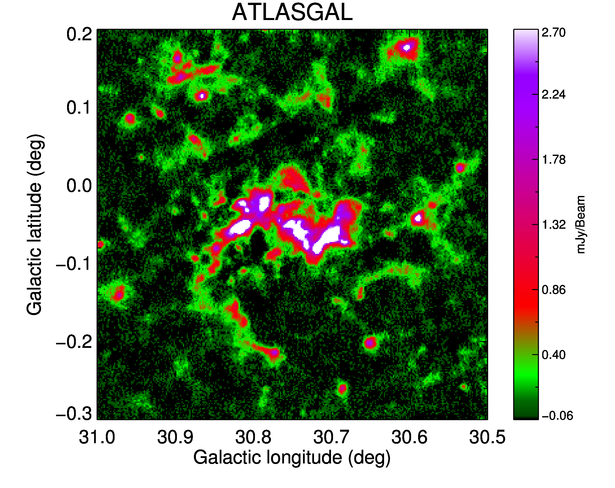} & \includegraphics[width=0.33\textwidth]{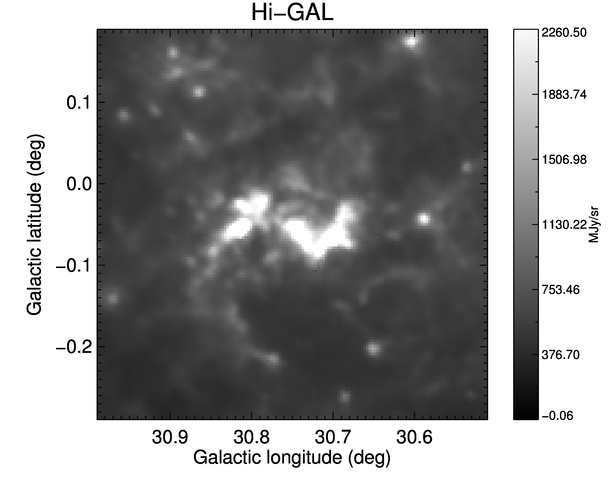} & \includegraphics[width=0.33\textwidth]{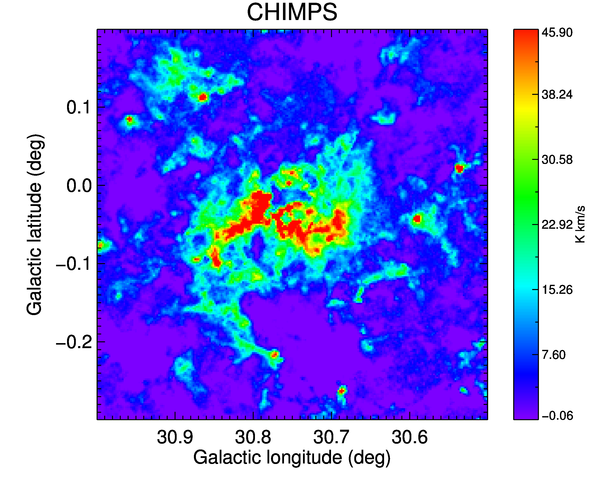} \\
\includegraphics[width=0.33\textwidth]{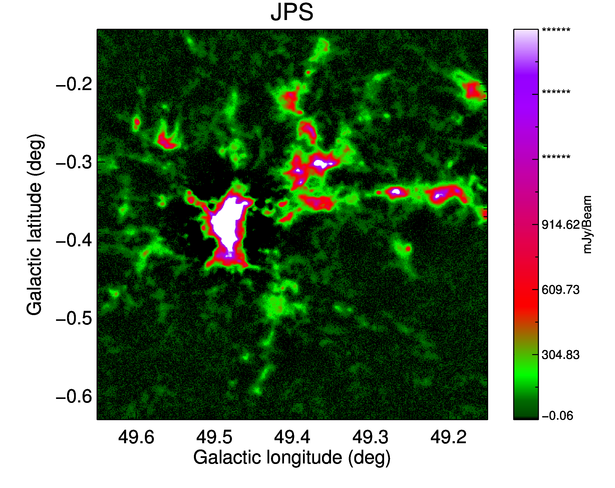} & \includegraphics[width=0.33\textwidth]{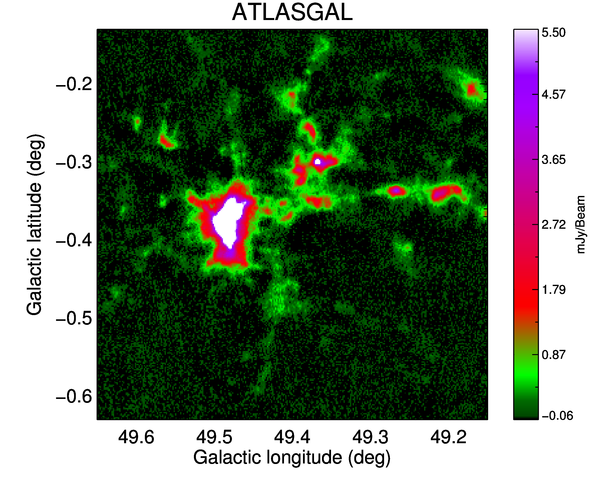} & \includegraphics[width=0.33\textwidth]{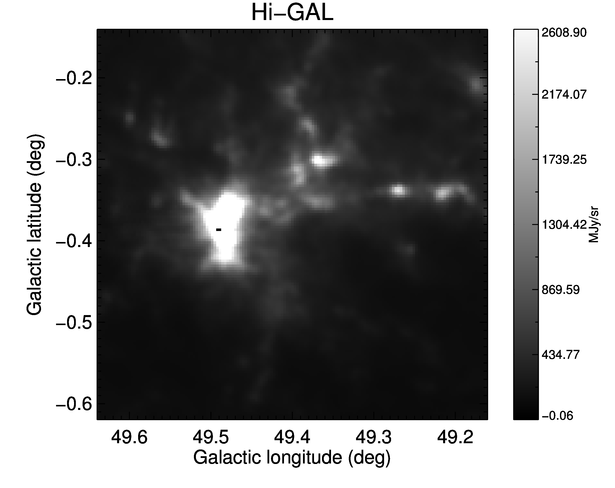} & \includegraphics[width=0.33\textwidth]{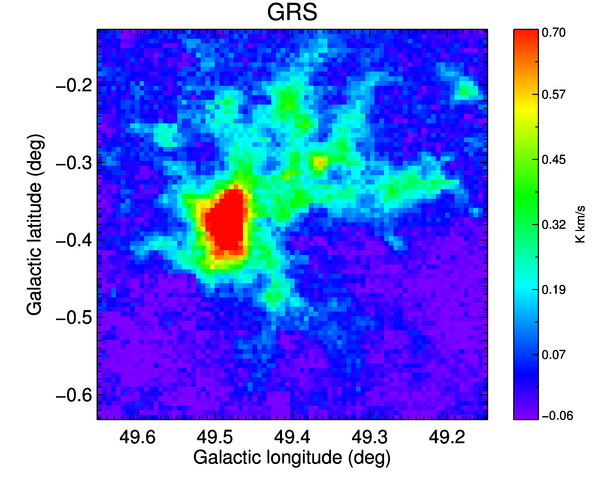} \\
\end{tabular}
\caption{Interesting sources within the JPS fields. The first row is the W39 \ion{H}{ii} region, the second row is the W43 star-forming region and the third row is the W51 star-forming region. The first column is the JPS data, with the second column containing the ATLASGAL data, the third column is the Hi-GAL 500-$\upmu$m data and the final column is an integrated molecular emission map, with W39 and W51 showing $^{13}$CO $J = 1-0$ emission from the GRS and W43 is $^{13}$CO $J = 3-2$ CHIMPS data.}
\label{closeups}
\end{center}
\end{figure}
\end{landscape}

Some of the more interesting sources in the JPS fields are highlighted in Fig.~\ref{closeups}. The JPS data (first column) are compared to the ATLASGAL and Hi-GAL 500-$\upmu$m images (second and third columns, respectively) as well as an integrated-intensity molecular map (final column). The molecular data are from two different surveys, due to data availability. The first and third rows, displaying the sources W39 and W51, use data from the GRS \citep{Jackson06} in the $^{13}$CO $J = 1-0$ transition. The source in the second row is W43, for which the molecular data are $^{13}$CO $J = 3-2$ from the CHIMPS survey \citep{Rigby16}.

The first row of Fig.~\ref{closeups} highlights the \ion{H}{ii} region W39, located at $\ell$\,=\,18.86$\degr$,\,$\emph{b}$\,=\,$-$0.48$\degr$. There are only a few studies of this region, with previous detections in radio recombination lines \citep{Lockman89} and infrared PAH emission \citep{Giard89}. The \ion{H}{ii} region morphology was identified in the GLIMPSE survey \citep{Churchwell06}. At an estimated distance of 4.5\,kpc, W39 has a diameter of $\sim$30\,pc \citep{Kerton13}. The size of this \ion{H}{ii} region implies it is relatively evolved, as the mean size of ultracompact \ion{H}{ii} regions is found to be significantly smaller \citep{Urquhart13}. The study of \citet{Kerton13} found evidence for potential sequential star formation, with the W39 \ion{H}{ii} region surrounded by smaller \ion{H}{ii} regions. Studies of YSOs around bubble structures has shown that a minor yet significant fraction (14-30 per cent) of Galactic star formation could be the result of triggering \citep{Thompson12,Kendrew12}, with overdensities of clumps also found \citep{Kendrew16}.

The W43 star-forming region, at $\ell$\,=\,30.75$\degr$,\,$\emph{b}$\,=\,$-$0.05$\degr$, is displayed in the second row of Fig.~\ref{closeups}. Along with W51, W43 is one of the two most striking regions within the JPS (Figs~\ref{dataimages} \&~\ref{dataimages2}). Thought to be located at the near end of the Long Bar of the Galaxy, at the tangent of the Scutum--Centaurus arm \citep{NguyenLuong11}, it has a distance of 5.5\,kpc \citep{Zhang14}. The presence of W43 at the end of the bar makes it a candidate to be an extreme star-forming region, similar to those regularly found at the ends of bars in external galaxies \citep*[e.g.][]{James09}.

Despite the prominence of W43 in the JPS, indicating an abundance of both gas \citep[e.g.][]{Rigby16} and dust, there is little evidence that the current star-formation efficiency of the region is presently enhanced, relative to the average in the Galactic disc. Previous studies have labelled the region as a `mini-starburst', implying a high star-formation efficiency \citep{Louvet14} and high future star-formation rate \citep{Motte03}. When placed in a Galactic context, however, there is no evidence that the existence of W43 is boosting the star-formation efficiency at the Galactocentric radius associated with the region \citep{Moore12,Eden15}, whereas other major regions, such as W49 and W51, do raise the average SFE at their corresponding radii. The molecular cloud associated with W43 does have an elevated clump formation efficiency but this value falls in the wings of a lognormal distribution. Hence, although extreme, W43's CFE is not abnormal, with clouds like it expected in any large enough sample \citep{Eden12}.

W51 (third row of Fig.~\ref{closeups}), located at $\ell$\,=\,49.40$\degr$,\,$\emph{b}$\,=\,$-$0.38$\degr$, does have some properties related to starburst conditions. For example, its high $L/M$ ratio of $\sim$ 13\,L$_{\sun}$/M$_{\sun}$ \citep{Harvey86,Kang10} is comparable to LIRGS and ULIRGS \citep[e.g.][]{Solomon97}. The W51 star-forming region has very efficient star formation \citep*{Kumar04} that has occurred recently \citep{Clark09}. In addition, \citet{Moore12} found an elevated SFE at the Galactocentric radius associated with W51, calculated from a distance of 5.4\,kpc \citep{Sato10}, due to an increased number of YSOs per unit molecular mass. The presence of this region in this Galactocentric radius bin has a significant influence on kpc-scale averages and is a candidate mini-starburst region.

The extra fidelity in the JPS data is most obvious in more diffuse areas of the maps, where the extended, faint emission can be seen that is present in the Hi-GAL images but not in the ATLASGAL data. Fig.~\ref{3surveys} shows an example region from the $\ell$\,=\,30$\degr$ region in the Hi-GAL 500-$\upmu$m, JPS and ATLASGAL data sets from top to bottom, respectively. There are no significant emission features in the Hi-GAL image that are not present in the JPS data, and vice versa, except for the large-scale extended background due to the diffuse Galactic-plane emission that is resolved out by the SCUBA-2 observing method. A quantitative analysis of the response to extended structure and flux calibrations relative to {\em Herschel} will be outlined in a future study (Tahani et al., in preparation).

\section{Global star-formation properties}

\subsection{Fraction of emission in compact sources}

The JPS compact-source catalogue (CSC) does not account for all of the emission in the JPSPR1 image data. The fraction extracted in each of the six JPS fields is listed in Table~\ref{sciresults}. The fraction is determined as the ratio of the emission within the catalogued sources to the total emission within the pixels in the SNR map above 5, the completeness threshold of the JPS. The ratios found for the individual regions are consistent with each other, with an average of 42 per cent, indicating that the fraction of emission contained within compact sources does not change significantly with Galactic longitude, at least in the inner Galaxy.

\begin{figure*}
\includegraphics[width=0.99\linewidth]{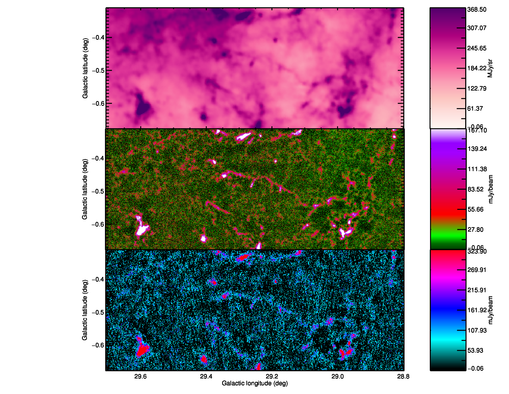} \\
\caption{A region of the $\ell$\,=\,30$\degr$ JPS field, as seen in the {\em Herschel} Hi-GAL 500-$\upmu$m, JPS 850-$\upmu$m and ATLASGAL 870-$\upmu$m data sets, in the top, middle and bottom panel, respectively. Whereas the large-scale diffuse emission seen by {\em Herschel} is filtered out in the ground-based JPS and ATLASGAL data, the fidelity of the JPS data to filamentary and compact structure can be clearly seen, as can the additional sensitivity compared to ATLASGAL.}
\label{3surveys}
\end{figure*}

The source-extraction process used to compile the CSC is not sensitive to filamentary structures, as seen by comparing the aspect ratios of sources here to those found in filaments \citep[e.g.][]{Schisano14}. As filaments are ubiquitous in the ISM, and JPS is not sensitive to diffuse structure, it is safe to assume that the $\sim$ 58 per cent of the detected emission not in compact sources is almost all associated with filamentary structures. Most of these are likely to be faint, as seen in Fig.~\ref{3surveys}.

The amount of emission detected by source-extraction methods is of the order of 50 per cent in other surveys, with the GRS reporting that 37 per cent of the mass is not detected in their cloud and clump catalogues \citep{Rathborne09}.

\subsection{Comparisons with molecular-line surveys}

Most, if not all, Galactic star formation occurs in molecular clouds. If the clumps in the JPS catalogue are expected to be forming stars, or to form them in the future, then JPS clumps are very likely to be within molecular clouds. By matching the JPS catalogue sources in $\ell$ and $\emph{b}$ space to the molecular clouds of the GRS \citep{Roman-Duval09} and the molecular clumps detected in CHIMPS (Rigby et al., in preparation), we can determine the percentage that falls on the same line of sight as these molecular structures. Three-dimensional matching, as done with the BGPS and the GRS in \citet{Eden12,Eden13} will form part of a further study, with distances determined for the JPS catalogue.

The GRS catalogue spans four of the JPS fields ($\ell$ = 20$\degr$ to 50$\degr$), whilst CHIMPS covers two ($\ell$ = 30$\degr$ and 40$\degr$). The fractions of JPS sources matched to molecular clouds and clumps within the coincident regions are shown in Table~\ref{sciresults}. The percentage of sources with a GRS molecular cloud in the same line of sight has a mean value of 98.2 per cent, meaning that almost all JPS sources which overlap with a GRS map have at least one potential molecular cloud with which they could be associated. Visual inspection of the GRS data for the remaining sources finds that there is also uncatalogued GRS emission along these lines of sight.

The CHIMPS clump catalogue \citep{Rigby16a} used to match the JPS sources is made up of emission extracted and presented in \citet{Rigby16}. CHIMPS traces denser molecular gas with the $^{13}$CO $J$ = $3-2$ transition than does GRS in $^{13}$CO $J$ = $1-0$. As a result, the emission in CHIMPS is not as ubiquitous as in GRS. Despite this, the fraction of JPS sources associated with CHIMPS clumps within the $\ell$ = 30$\degr$ field is consistent with the GRS fraction. The match rate in $\ell$ = 40$\degr$ is a little lower than for GRS, but the difference is less than 2$\sigma$ and so not significant. This suggests that random positional matches between unrelated structures along the line of sight are relatively uncommon.

The percentage of sources found to be associated with a molecular cloud is much higher than that found in \citet{Eden12,Eden13}, who found $\sim$ 80 per cent of sources associated with GRS clouds. The lower percentages found in those studies are due to the added dimension of velocity, not taken into account here, with sources found to be associated with real structure not catalogued by the GRS (see \citealt{Eden12} for a full discussion. The combination of these results implies that the chance association rate is $\sim$10-15 per cent.

\subsection{Properties of star-forming clumps}

Without consistent distance information for the compact sources in the current study, we shall not present any properties of the sample that involve masses or luminosities.  Other properties that do not require distances can, however, be examined.

The Hi-GAL survey traces the YSO content of the Galaxy, with the detection of a 70-$\upmu$m point source taken as reliable evidence of the presence of a protostar \citep[e.g.][]{Dunham08,Ragan12,Veneziani13}. By positionally matching sources in the Hi-GAL band-merged catalogue that contain a 70-$\upmu$m point source (Elia et al., in preparation), with the JPS catalogue within 14.4 arcsec, we can determine which JPS sources host a protostar. These matches resulted in 3,056 70-$\upmu$m sources associated with 2,946 JPS clumps; therefore, $\sim$ 38 per cent of all JPS sources are currently star-forming. \citep{Svoboda16} found that $\sim$ 46 per cent of BGPS sources were coincident with a 70-$\upmu$m point source. In comparison, 51 per cent of the objects found in the W43 star-forming region in the \citet{Moore15} study are currently forming stars.

\begin{figure*}
\begin{tabular}{ll}
\includegraphics[scale=0.50]{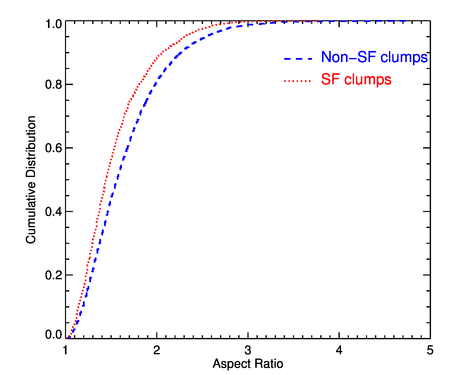} & \includegraphics[scale=0.50]{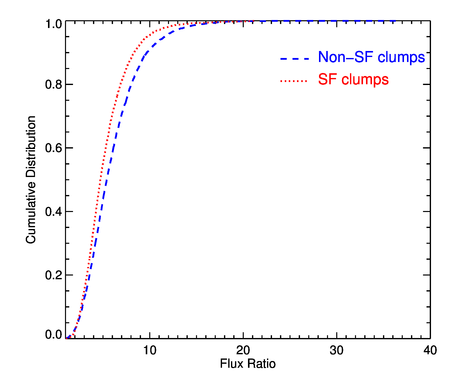}\\
\end{tabular}
\caption{The cumulative distributions of the aspect ratios and integrated-to-peak fluxes, in the left and right panels, respectively. The JPS clumps with a 70\,$\upmu$m source are represented by the red dotted line, whilst the entire JPS population is represented by the blue dashed line.}
\label{SFratios}
\end{figure*}

The compactness of the clumps, measured by their aspect ratio and integrated-to-peak flux ratio, can also be investigated. The integrated-to-peak flux ratio, also known as the compactness factor or $Y$-factor, estimates how centrally condensed the emission is, with a low value meaning that the emission is more centrally condensed in the clump. In Fig.~\ref{SFratios}, we present the cumulative distributions of the aspect ratios and the integrated-to-peak flux ratios for both the star-forming clumps identified above and clumps not associated with a YSO. The clumps with a 70-$\upmu$m source are noticeably different, in both distributions, to the complete JPS sample, having a skew to more compact sources. The mean and median values also reflect this difference. For the aspect ratio, the mean and median values are 1.54\,$\pm$0.01 and 1.44 for clumps with an associated 70-$\upmu$m source, respectively, compared to 1.67\,$\pm$0.01 and 1.56 for the rest of JPS sample. The equivalent values from the integrated-to-peak flux ratio distributions are 5.24\,$\pm$0.05 and 4.72, and 5.98\,$\pm$0.04 and 5.40, respectively. K--S tests of the samples indicate that we can conclude that in both cases, the samples can be considered to be drawn from different populations.

The combination of these results indicates that a star-forming clump is more centrally condensed than those that are not. Either the clumps without the star-formation indicator need to be initially centrally condensed or they become so soon after they begin to form stars. This result is consistent with that of \citet{Urquhart14a} who found that a sample of ATLASGAL clumps associated with masers, \ion{H}{ii} regions, and YSOs is more centrally concentrated than the one of clumps which do not host a star-formation indicator. This is also seen in the nearby star-forming region Orion B in the JCMT Gould Belt Survey \citet{Kirk16}.

\begin{table}
\begin{center}
\caption{The fractions of the total JPS emission contained in the CSC, and of numbers of CSC sources associated with GRS molecular clouds and CHIMPS clumps, in each of the six JPS fields.}
\label{sciresults}
\begin{tabular}{lccc}\hline
Field & Fraction of  & Fraction associated & Fraction associated\\
 & JPS emission & with GRS clouds & with CHIMPS clumps\\
\hline
$\ell$ = 10$\degr$ & 0.41 $\pm$ 0.03 & ... & ...\\
$\ell$ = 20$\degr$ & 0.46 $\pm$ 0.04 & 0.991 $\pm$ 0.035 & ...\\
$\ell$ = 30$\degr$ & 0.41 $\pm$ 0.03 & 0.977 $\pm$ 0.030 & 0.93 $\pm$ 0.03\\
$\ell$ = 40$\degr$ & 0.54 $\pm$ 0.06 & 0.989 $\pm$ 0.046 & 0.87 $\pm$ 0.05\\
$\ell$ = 50$\degr$ & 0.33 $\pm$ 0.08 & 0.964 $\pm$ 0.041 & ...\\
$\ell$ = 60$\degr$ & 0.46 $\pm$ 0.07 & ... & ...\\
\hline
Total & 0.42 $\pm$ 0.05 & 0.982 $\pm$ 0.018 & 0.91 $\pm$ 0.03\\
\hline
\end{tabular}
\end{center}
\end{table}

\section{Summary}

The first public data release of the JCMT Plane Survey is presented, including 850-$\upmu$m continuum images and a compact source catalogue. The data are publicly available and can be downloaded from the CANFAR archive. The image data reach an average pixel-to-pixel noise of 7.19 mJy\,beam$^{-1}$, when smoothed over the beam.

The compact-source extraction, carried out using the {\sc FellWalker} algorithm, resulted in a catalogue of 7,813 sources above a 5-$\upsigma$ threshold. 38\,$\pm$\,1 per cent of these are associated with a {\em Herschel} 70-$\upmu$m source and so can be considered to be a star-forming region.  The JPSPR1 compact catalogue sources contribute 42\,$\pm$\,5 per cent of the total emission in the images. The remainder of the 850-$\upmu$m emission in the image data is assumed to arise in filamentary structures.

Completeness testing within the six individual fields of the JPS finds that a 95 per-cent completeness limit is reached at 5\,$\upsigma$ in the fields centred at $\ell$\,=\, 10$\degr$, 20$\degr$, 50$\degr$ and 60$\degr$ ($\sim$\,140 mJy\,beam$^{-1}$), with the $\ell$\,=\,30$\degr$ and 40$\degr$ fields reaching this completeness at 7.5\,$\upsigma$ and 6.5\,$\upsigma$, respectively, corresponding to 224 and 181 mJy\,beam$^{-1}$, respectively. The higher completeness thresholds indicate that the confusion limit has been reached in these two fields.  

The integrated flux densities of JPS compact sources are found to be systematically lower than those of positionally matched ATLASGAL sources. This is the result of the improved spatial resolution of the JPS data, which tends to reveal substructure in the ATLASGAL sources. Detected structure often depends on the spatial resolution of the data and care should be taken to select the data most appropriate to the intended science.
The distributions of the flux densities of sources in each survey show that the JPS is around 10 times more sensitive than ATLASGAL, with the 95 per cent completeness limits estimated to be 0.04\,Jy\,beam$^{-1}$ and 0.3\,Jy for the peak and integrated flux densities, respectively.

The JPSPR1 compact-source catalogue and images were also compared to other surveys of the Galactic Plane. Positionally matching the compact source catalogue to the molecular cloud catalogues of the GRS and CHIMPS surveys in the overlap regions reveals that 98\,$\pm$\,2 per cent of JPS sources are associated with GRS-catalogued $^{13}$CO\,$J\,=\,1-0$ emission and 91\,$\pm$\,3 per cent are associated with $^{13}$CO\,$J\,=\,3-2$ emission tracing denser gas detected by the CHIMPS survey.

The star-forming fraction of the JPS sources was found to be 38\,$\pm$\,1 per cent, after positionally matching the JPSPR1 catalogue with the band-merged catalogue of the Hi-GAL survey. The compactness of the JPS sources, measured from both the aspect ratio and the ratio of the integrated and peak fluxes shows that those sources associated with a potential YSO are more compact than those of the rest of the sample.

\section*{Acknowledgements}

DJE is supported by a STFC postdoctoral grant (ST/M000966/1). The JCMT has historically been operated by the Joint Astronomy Centre on behalf of the Science and Technology Facilities Council of the United Kingdom, the National Research Council of Canada and the Netherlands Organization for Scientific Research. The data presented in this paper were taken under JCMT observing programme MJLSJ02. Additional funds for the construction of SCUBA-2 were provided by the Canada Foundation for Innovation. This research has made use of NASA's Astrophysics Data System. The Starlink software \citep{Currie14} is currently supported by the East Asian Observatory. GJW gratefully thanks the Leverhulme Trust for the support of an Emeritus Fellowship.

\begin{landscape}
\begin{table}
\begin{center}
\caption{The JPSPR1 compact source catalogue. The columns are as follows: (1) JPS catalogue source name; (2) IAU source identifier; (3) and (4) Galactic coordinates of the peak flux position within the JPS source; (5) and (6) Galactic coordinates of the central point; (7--9) semi-major, semi-minor and position angle, measured anticlockwise from the Galactic north, of the ellipse fit to the shape of the JPS source; (10) effective radius of source, calculated by $\sqrt{(A/\pi)}$, where $A$ is the area of the source above the detection threshold; (11--12) peak flux density, in units of Jy\,beam$^{-1}$, and associated uncertainty; (13--14) integrated flux, in units of Jy, and associated uncertainty and (15) signal-to-noise ratio (SNR) of the source, calculated from the uncorrected peak flux density and the unsmoothed $\upsigma_{\rmn{rms}}$ from Table.~\ref{rmsvalues}.}
\label{sourcecatalogue}
\begin{tabular}{llcccccccccccccc}\hline
Name & IAU Designation & $\ell_{\rmn{peak}}$ & $\emph{b}_{\rmn{peak}}$ & $\ell_{\rmn{cen}}$ & $\emph{b}_{\rmn{cen}}$ & $\upsigma_{\rmn{maj}}$ & $\upsigma_{\rmn{min}}$ & PA & $R_{\rmn{eff}}$ & $S_{\rmn{peak}}$ & $\Delta$$S_{\rmn{peak}}$ & $S_{\rmn{int}}$ & $\Delta$$S_{\rmn{int}}$ & SNR\\
& & ($^{\circ}$) & ($^{\circ}$) & ($^{\circ}$) & ($^{\circ}$) & ($\prime\prime$) & ($\prime\prime$) & ($^{\circ}$) & ($\prime\prime$) & (Jy\,beam$^{-1}$) & (Jy\,beam$^{-1}$) & (Jy) & (Jy) & \\
(1) & (2) & (3) & (4) & (5) & (6) & (7) & (8) & (9) & (10) & (11) & (12) & (13) & (14) & (15) \\
\hline
JPSG006.687$-$00.294	&	JCMTLSPJ180152.1$-$231918	&	6.687	&	$-$0.294	&	6.688	&	$-$0.292	&	16	&	5	&	120	&	17	&	0.484	&	0.024	&	1.166	&	0.124	&	17.04	\\
JPSG006.750$-$00.341	&	JCMTLSPJ180210.9$-$231723	&	6.750	&	$-$0.341	&	6.751	&	$-$0.338	&	12	&	5	&	144	&	13	&	0.179	&	0.009	&	0.346	&	0.061	&	8.61	\\
JPSG006.760$-$00.284	&	JCMTLSPJ180159.3$-$231511	&	6.760	&	$-$0.284	&	6.759	&	$-$0.281	&	8	&	6	&	143	&	14	&	0.151	&	0.008	&	0.313	&	0.059	&	7.86	\\
JPSG006.770$-$00.269	&	JCMTLSPJ180157.2$-$231413	&	6.770	&	$-$0.269	&	6.767	&	$-$0.267	&	10	&	4	&	192	&	12	&	0.168	&	0.009	&	0.346	&	0.063	&	8.31	\\
JPSG006.777$-$00.263	&	JCMTLSPJ180156.7$-$231342	&	6.777	&	$-$0.263	&	6.774	&	$-$0.264	&	7	&	6	&	121	&	13	&	0.229	&	0.012	&	0.473	&	0.061	&	9.98	\\
JPSG006.786$-$00.374	&	JCMTLSPJ180223.0$-$231630	&	6.786	&	$-$0.374	&	6.785	&	$-$0.374	&	10	&	6	&	156	&	17	&	0.186	&	0.010	&	0.625	&	0.057	&	8.81	\\
JPSG006.796$-$00.258	&	JCMTLSPJ180158.0$-$231233	&	6.796	&	$-$0.258	&	6.795	&	$-$0.258	&	23	&	16	&	169	&	50	&	3.720	&	0.198	&	16.051	&	0.804	&	119.78	\\
JPSG006.811$-$00.395	&	JCMTLSPJ180231.2$-$231550	&	6.811	&	$-$0.395	&	6.812	&	$-$0.395	&	6	&	6	&	180	&	13	&	0.070	&	0.004	&	0.244	&	0.045	&	5.68	\\
JPSG006.826$-$00.225	&	JCMTLSPJ180154.3$-$230960	&	6.826	&	$-$0.225	&	6.825	&	$-$0.221	&	11	&	9	&	116	&	19	&	0.207	&	0.011	&	0.675	&	0.057	&	9.38	\\
JPSG006.829$-$00.114	&	JCMTLSPJ180129.5$-$230635	&	6.829	&	$-$0.114	&	6.829	&	$-$0.115	&	9	&	6	&	269	&	15	&	0.202	&	0.011	&	0.434	&	0.063	&	9.23	\\
JPSG006.865$-$00.241	&	JCMTLSPJ180202.9$-$230828	&	6.865	&	$-$0.241	&	6.865	&	$-$0.243	&	7	&	4	&	131	&	11	&	0.063	&	0.004	&	0.189	&	0.047	&	5.51	\\
JPSG006.868$-$00.434	&	JCMTLSPJ180247.2$-$231401	&	6.868	&	$-$0.434	&	6.867	&	$-$0.432	&	16	&	9	&	265	&	25	&	0.294	&	0.017	&	0.987	&	0.057	&	11.76	\\
JPSG006.885$-$00.022	&	JCMTLSPJ180115.9$-$230053	&	6.885	&	$-$0.022	&	6.885	&	$-$0.021	&	9	&	7	&	113	&	17	&	0.187	&	0.010	&	0.590	&	0.073	&	8.83	\\
JPSG006.905$-$00.225	&	JCMTLSPJ180204.4$-$230555	&	6.905	&	$-$0.225	&	6.899	&	$-$0.228	&	29	&	20	&	125	&	51	&	0.502	&	0.028	&	5.877	&	0.295	&	17.57	\\
JPSG006.905$-$00.453	&	JCMTLSPJ180256.4$-$231237	&	6.905	&	$-$0.453	&	6.902	&	$-$0.452	&	13	&	5	&	258	&	17	&	0.052	&	0.003	&	0.318	&	0.034	&	5.21	\\
JPSG006.912$-$00.266	&	JCMTLSPJ180214.7$-$230644	&	6.912	&	$-$0.266	&	6.913	&	$-$0.265	&	11	&	9	&	209	&	23	&	0.371	&	0.022	&	1.200	&	0.067	&	13.89	\\
JPSG006.912$-$00.276	&	JCMTLSPJ180217.1$-$230703	&	6.912	&	$-$0.276	&	6.912	&	$-$0.276	&	10	&	8	&	267	&	22	&	0.320	&	0.019	&	0.916	&	0.054	&	12.48	\\
JPSG006.915$-$00.226	&	JCMTLSPJ180206.0$-$230522	&	6.915	&	$-$0.226	&	6.920	&	$-$0.225	&	23	&	17	&	147	&	51	&	1.614	&	0.093	&	10.671	&	0.534	&	51.97	\\
JPSG006.919$-$00.561	&	JCMTLSPJ180322.7$-$231507	&	6.919	&	$-$0.561	&	6.919	&	$-$0.561	&	12	&	8	&	254	&	20	&	0.160	&	0.009	&	0.574	&	0.040	&	8.11	\\
JPSG006.923$-$00.252	&	JCMTLSPJ180213.0$-$230546	&	6.923	&	$-$0.252	&	6.921	&	$-$0.253	&	16	&	11	&	131	&	34	&	1.110	&	0.065	&	3.906	&	0.197	&	35.74	\\
JPSG006.924$-$00.391	&	JCMTLSPJ180244.7$-$230950	&	6.924	&	$-$0.391	&	6.924	&	$-$0.389	&	7	&	4	&	251	&	11	&	0.069	&	0.004	&	0.130	&	0.037	&	5.65	\\
JPSG006.926$+$00.038	&	JCMTLSPJ180107.8$-$225660	&	6.926	&	0.038	&	6.925	&	0.038	&	8	&	5	&	210	&	13	&	0.242	&	0.012	&	0.471	&	0.097	&	10.32	\\
JPSG006.927$-$00.570	&	JCMTLSPJ180325.8$-$231457	&	6.927	&	$-$0.570	&	6.928	&	$-$0.568	&	12	&	11	&	166	&	24	&	0.111	&	0.007	&	0.827	&	0.049	&	6.78	\\
JPSG006.933$-$00.206	&	JCMTLSPJ180203.8$-$230352	&	6.933	&	$-$0.206	&	6.932	&	$-$0.205	&	9	&	7	&	101	&	17	&	0.172	&	0.010	&	0.504	&	0.041	&	8.42	\\
JPSG006.942$-$00.282	&	JCMTLSPJ180222.3$-$230539	&	6.942	&	$-$0.282	&	6.944	&	$-$0.283	&	17	&	10	&	187	&	28	&	0.136	&	0.008	&	1.182	&	0.065	&	7.44	\\
JPSG006.943$-$00.294	&	JCMTLSPJ180225.1$-$230558	&	6.943	&	$-$0.294	&	6.940	&	$-$0.294	&	13	&	6	&	167	&	17	&	0.075	&	0.004	&	0.348	&	0.035	&	5.81	\\
JPSG006.958$-$00.261	&	JCMTLSPJ180219.6$-$230413	&	6.958	&	$-$0.261	&	6.958	&	$-$0.262	&	10	&	6	&	120	&	18	&	0.161	&	0.010	&	0.538	&	0.039	&	8.12	\\
JPSG006.963$-$00.311	&	JCMTLSPJ180231.4$-$230525	&	6.963	&	$-$0.311	&	6.959	&	$-$0.311	&	18	&	12	&	232	&	25	&	0.090	&	0.005	&	0.760	&	0.046	&	6.21	\\
JPSG006.965$-$00.286	&	JCMTLSPJ180226.1$-$230432	&	6.965	&	$-$0.286	&	6.964	&	$-$0.284	&	20	&	9	&	230	&	29	&	0.234	&	0.014	&	1.441	&	0.076	&	10.11	\\
JPSG006.975$-$00.179	&	JCMTLSPJ180203.2$-$230052	&	6.975	&	$-$0.179	&	6.978	&	$-$0.182	&	12	&	7	&	250	&	17	&	0.053	&	0.003	&	0.281	&	0.032	&	5.23	\\
JPSG006.982$-$00.288	&	JCMTLSPJ180228.8$-$230345	&	6.982	&	$-$0.288	&	6.979	&	$-$0.291	&	15	&	10	&	111	&	26	&	0.534	&	0.032	&	1.643	&	0.086	&	18.47	\\
JPSG006.984$-$00.226	&	JCMTLSPJ180215.1$-$230147	&	6.984	&	$-$0.226	&	6.984	&	$-$0.223	&	15	&	6	&	140	&	19	&	0.080	&	0.005	&	0.452	&	0.036	&	5.95	\\
JPSG006.984$-$00.259	&	JCMTLSPJ180222.4$-$230248	&	6.984	&	$-$0.259	&	6.982	&	$-$0.261	&	15	&	7	&	117	&	21	&	0.085	&	0.005	&	0.581	&	0.039	&	6.08	\\
JPSG006.994$-$00.232	&	JCMTLSPJ180217.7$-$230126	&	6.994	&	$-$0.232	&	6.996	&	$-$0.233	&	6	&	4	&	194	&	10	&	0.073	&	0.004	&	0.205	&	0.038	&	5.75	\\
JPSG006.996$-$00.216	&	JCMTLSPJ180214.4$-$230053	&	6.996	&	$-$0.216	&	6.995	&	$-$0.215	&	13	&	8	&	259	&	20	&	0.107	&	0.006	&	0.608	&	0.041	&	6.68	\\
JPSG006.999$-$00.243	&	JCMTLSPJ180220.7$-$230133	&	6.999	&	$-$0.243	&	6.992	&	$-$0.246	&	25	&	14	&	153	&	39	&	0.188	&	0.011	&	2.645	&	0.134	&	8.85	\\
JPSG006.999$-$00.575	&	JCMTLSPJ180336.0$-$231120	&	6.999	&	$-$0.575	&	7.002	&	$-$0.573	&	7	&	5	&	166	&	13	&	0.054	&	0.003	&	0.225	&	0.034	&	5.26	\\
JPSG007.000$-$00.365	&	JCMTLSPJ180248.5$-$230503	&	7.000	&	$-$0.365	&	7.001	&	$-$0.366	&	9	&	7	&	160	&	17	&	0.088	&	0.005	&	0.363	&	0.033	&	6.17	\\
\hline
\multicolumn{15}{l}{$\emph{Note:}$ Only a small portion of the catalogue is shown here. The entire catalogue is downloadable from the CANFAR archive and is also available in the Supporting Information.}\\
\end{tabular}
\end{center}
\end{table}
\end{landscape}

\bibliographystyle{mnras}
\bibliography{JPS_source_ref}

\appendix

\section{SMURF:MAKEMAP Parameters}
\label{smurfappend}

The following {\sc smurf:makemap} configuration parameters were used in the initial JPS data reduction process (in addition to a pixel size of 3 arcsec):\\

\noindent \texttt{\string^dimmconfig.lis\\
{\sc numiter} = -100 \\
{\sc flt.filt\_edge\_largescale} = 480 \\
{\sc flagslow} = 300 \\
{\sc maptol} = 0.01 \\
{\sc noi.box\_size} = -15 \\
{\sc noi.box\_type} = 1 \\
{\sc ast.zero\_mask} = 0 \\
{\sc ast.zero\_snr} = 3 \\
{\sc ast.zero\_snrlo} = 2 \\
{\sc ast.zero\_notlast} = 1 \\
{\sc flt.filt\_edge\_largescale\_last} = 100 \\
{\sc flt.ring\_box1} = 0.5 \\
{\sc flt.filt\_order} = 4 \\
{\sc com.sig\_limit} = 5 \\
}

The following {\sc smurf:makemap} configuration parameters were used in the JPS data reduction process with the external mask provided by combining the observations reduced in the manner above, setting regions of emission to one and background regions to zero (in addition to a pixel size of 3 arcsec):\\

\noindent \texttt{\string^dimmconfig.lis\\
{\sc numiter} = -100 \\
{\sc flt.filt\_edge\_largescale} = 480 \\
{\sc flagslow} = 300 \\
{\sc maptol} = 0.01 \\
{\sc noi.box\_size} = -15 \\
{\sc noi.box\_type} = 1 \\
{\sc ast.zero\_mask} = 1 \\
{\sc ast.zero\_snr} = 0 \\
{\sc ast.zero\_notlast} = 1 \\
{\sc flt.filt\_edge\_largescale\_last} = 100 \\
{\sc flt.ring\_box1} = 0.5 \\
{\sc flt.filt\_order} = 4 \\
{\sc com.sig\_limit} = 5 \\
}

\section{Histograms of pixel fluxes and noise}
\label{pixhistoappend}

\begin{figure*}
\begin{tabular}{lll}
\includegraphics[scale=0.32]{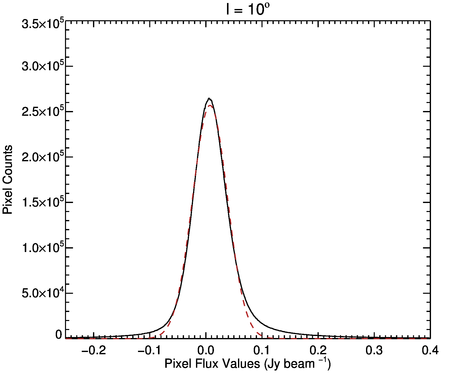} & \includegraphics[scale=0.32]{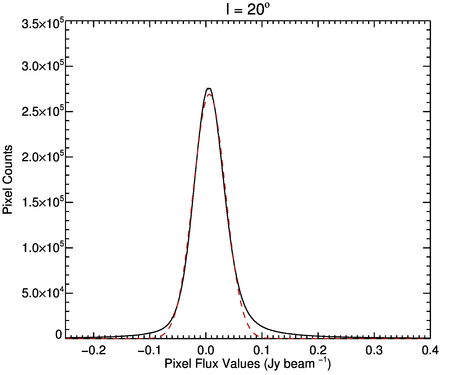} & \includegraphics[scale=0.32]{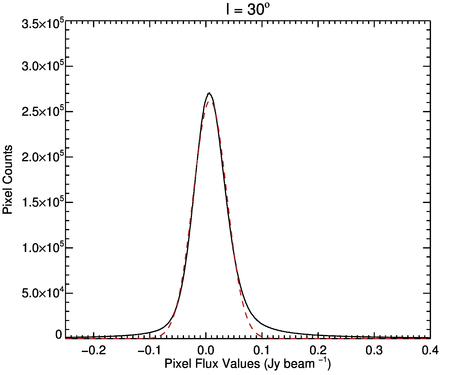}\\
\includegraphics[scale=0.32]{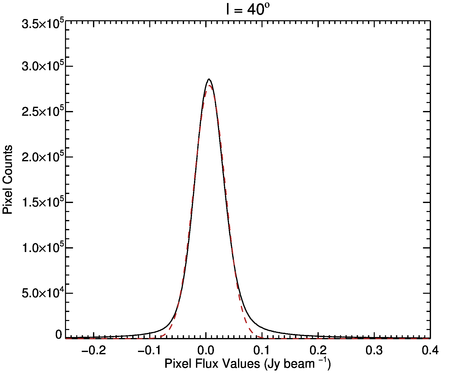} & \includegraphics[scale=0.32]{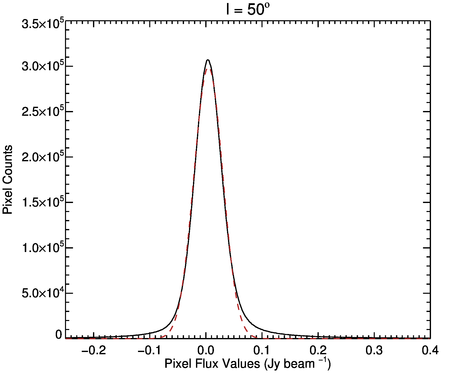} & \includegraphics[scale=0.32]{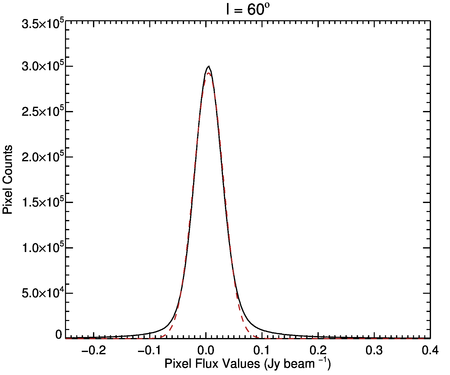}\\
\end{tabular}
\caption{The distributions of all pixel values in the six JPS fields are displayed in the black histograms, with the result of a Gaussian fit shown with a dashed red line.}
\label{pixhisto}
\end{figure*}

\begin{figure*}
\begin{tabular}{lll}
\includegraphics[scale=0.32]{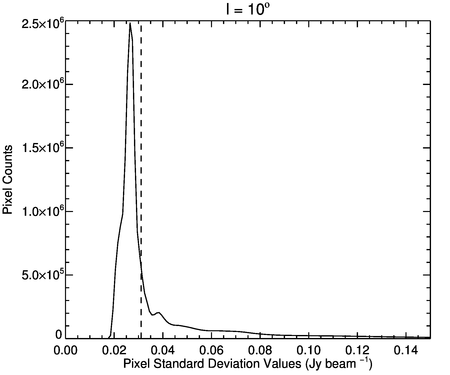} & \includegraphics[scale=0.32]{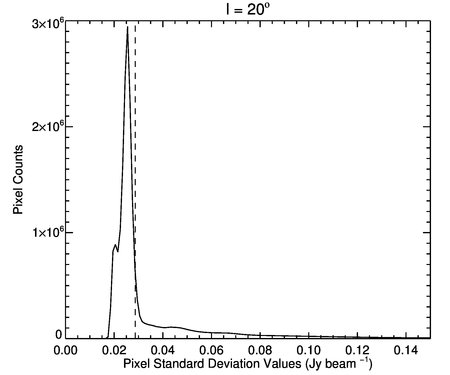} & \includegraphics[scale=0.32]{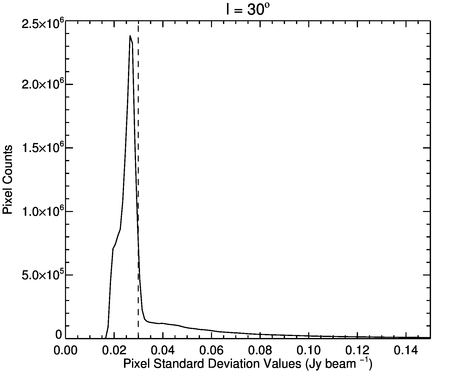}\\
\includegraphics[scale=0.32]{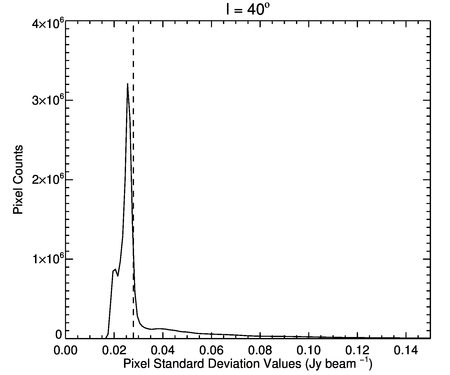} & \includegraphics[scale=0.32]{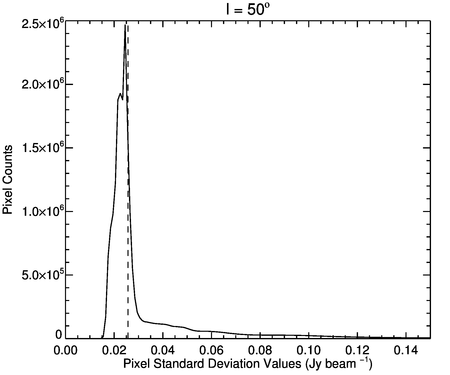} & \includegraphics[scale=0.32]{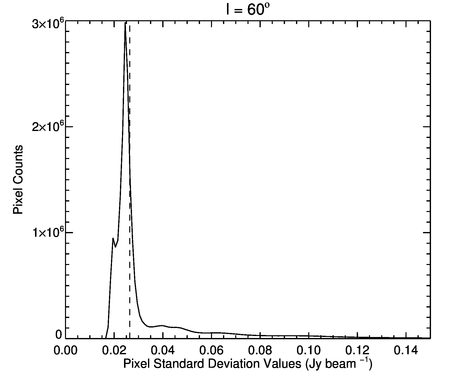}\\
\end{tabular}
\caption{Histograms of the noise values in each field of the JPS data, calculated from taking the square root of the variance arrays. The dashed vertical line represents the result of the fit from Fig.\ref{pixhisto}.}
\label{noisehisto}
\end{figure*}

Two methods for calculating the noise in each field are to plot the histogram of both the data and the square root of the variance maps, with these histograms shown in Fig.~\ref{pixhisto} and Fig.~\ref{noisehisto}. The calculation from the data histogram consists of fitting a Gaussian to the data and taking the width as an estimate of the noise. Using the variance data, the peak of the square root histogram is an estimate of the noise.

\section{Variance images}
\label{varianceappend}

\begin{figure*}
\begin{tabular}{l}
\includegraphics[width=\linewidth]{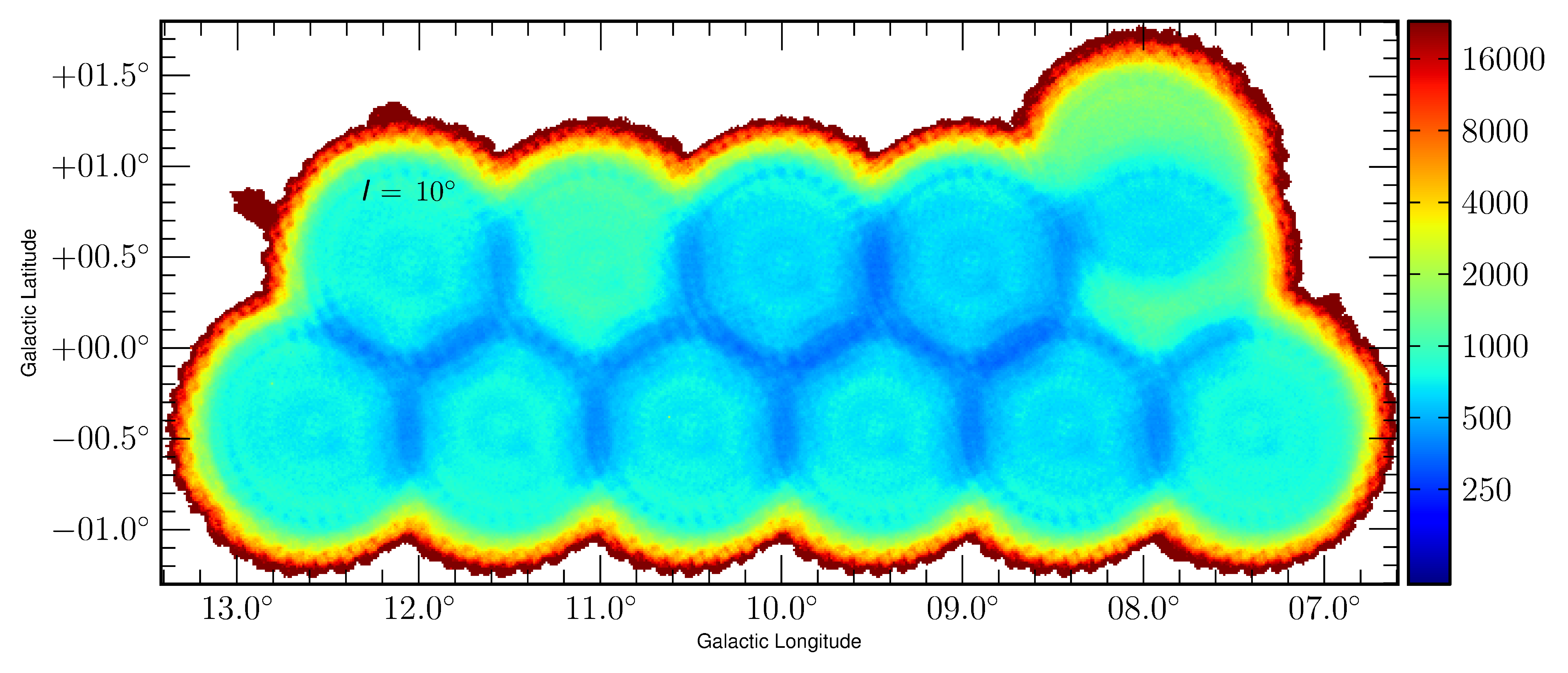}\\
\includegraphics[width=\linewidth]{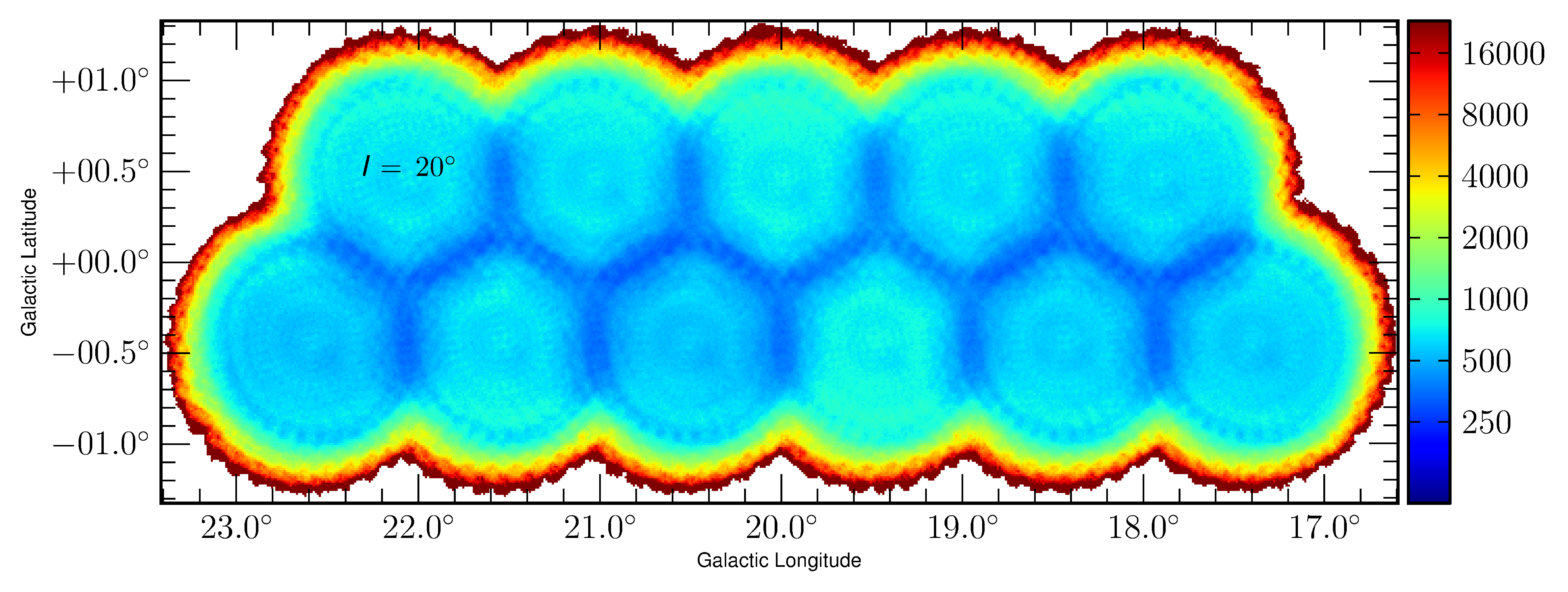}\\
\includegraphics[width=\linewidth]{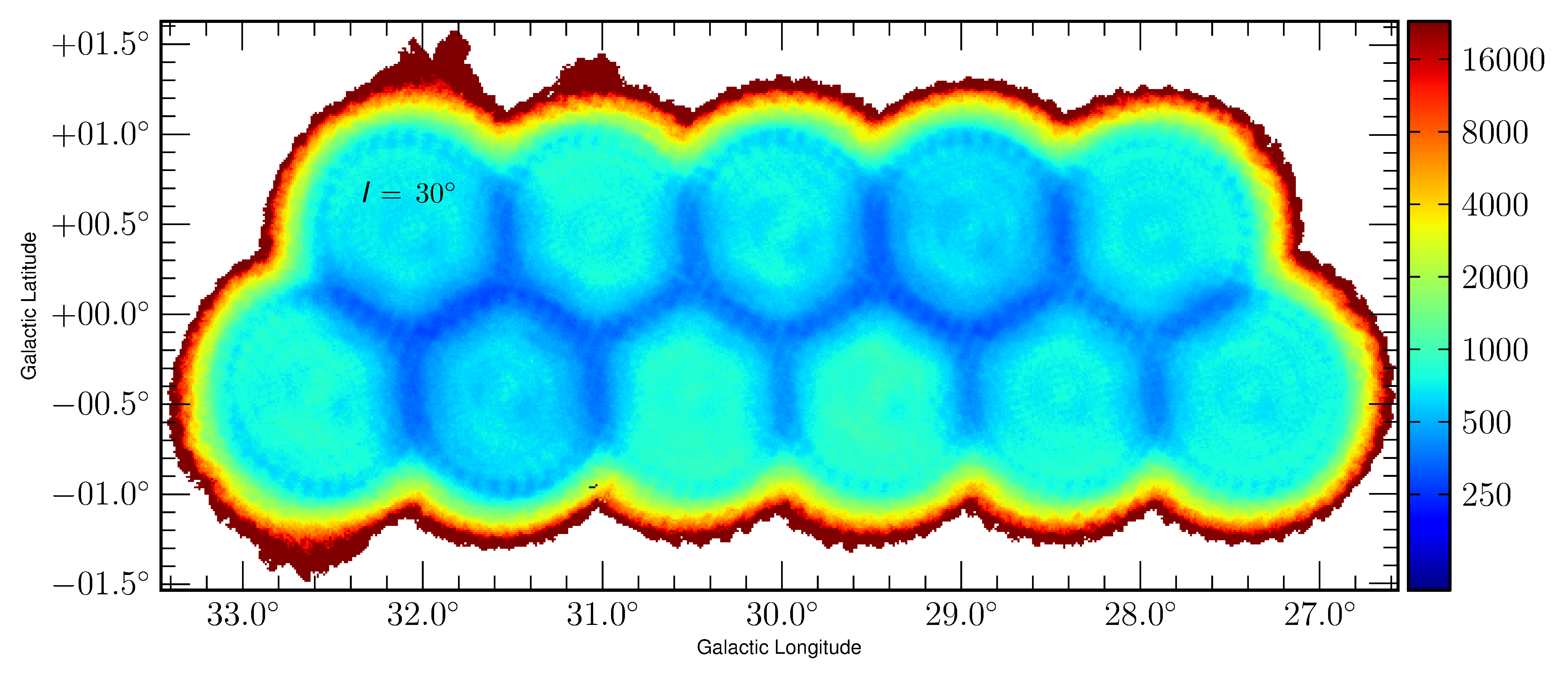}\\
\end{tabular}
\caption{Variance images for the $\ell$\,=\,10$\degr$, 20$\degr$ and 30$\degr$ fields, with the intensity scale in units of (mJy\,beam$^{-1}$)$^{2}$.}
\label{varianceimages}
\end{figure*}

\begin{figure*}
\begin{tabular}{l}
\includegraphics[width=\linewidth]{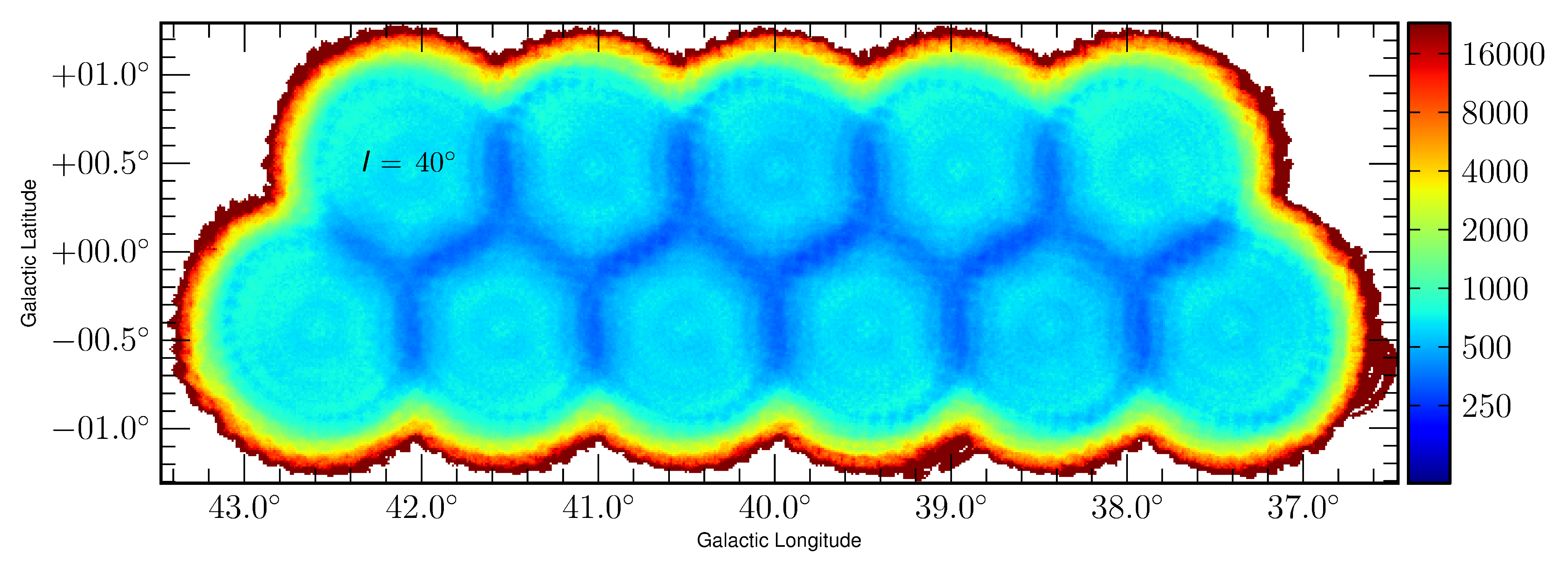}\\
\includegraphics[width=\linewidth]{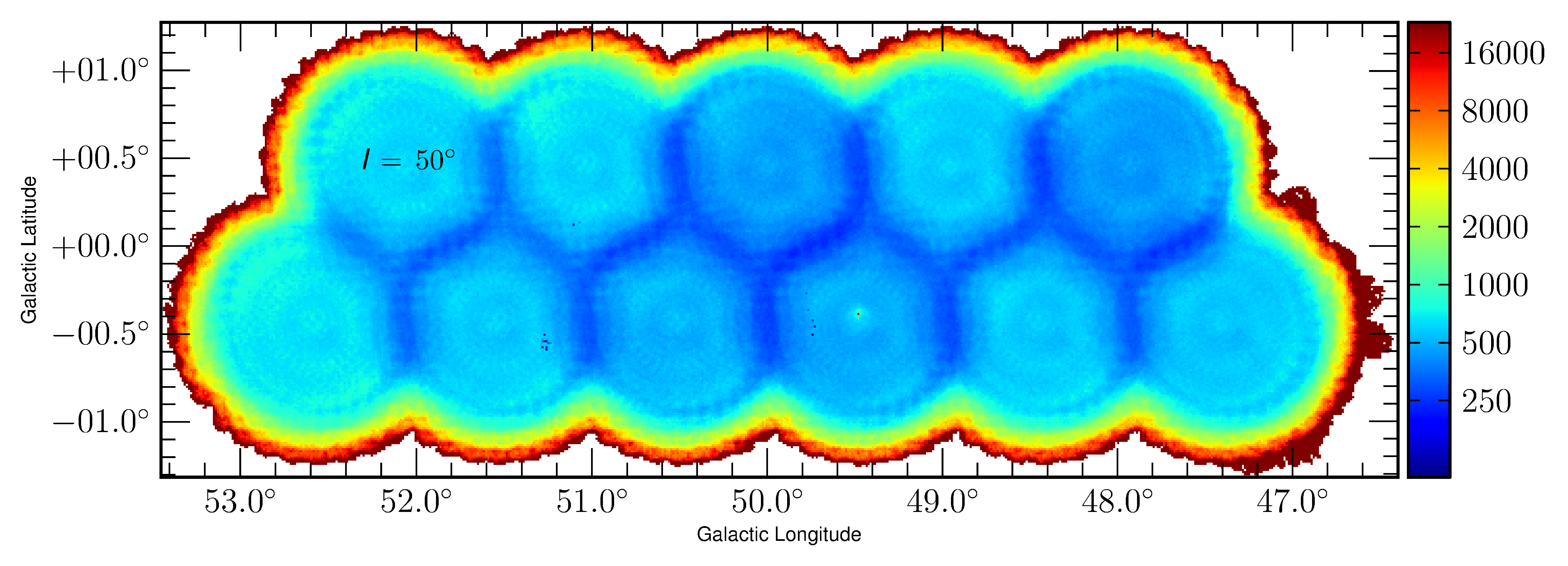}\\
\includegraphics[width=\linewidth]{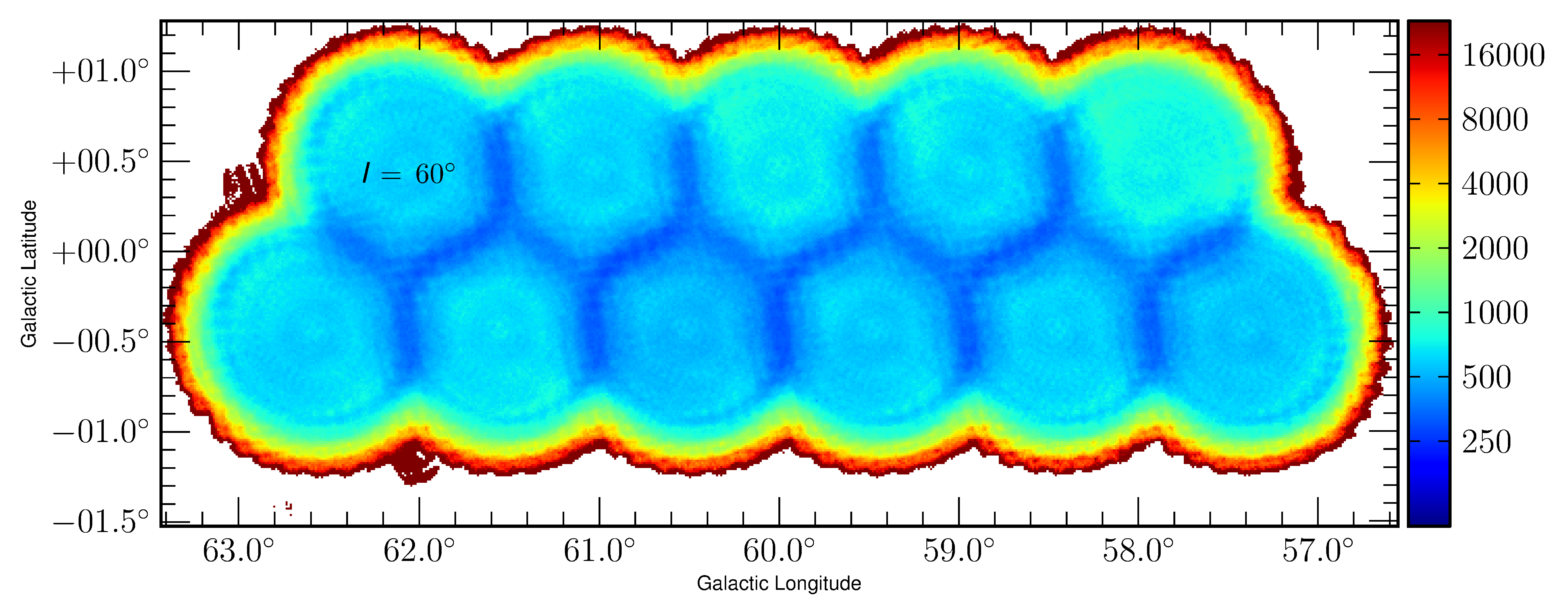}\\
\end{tabular}
\caption{Variance images for the $\ell$\,=\,40$\degr$, 50$\degr$ and 60$\degr$ fields, with the intensity scale in units of (mJy\,beam$^{-1}$)$^{2}$.}
\label{varianceimages2}
\end{figure*}

The variance images, produced by the data-processing and reduction software, corresponding to each JPS field can be found in Figs.~\ref{varianceimages} and ~\ref{varianceimages2}.

\section{Fellwalker Configuration Parameters}
\label{fellparamsappend}

The following {\sc fellwalker} configuration parameters were used in the source extraction process for the JPS compact source catalogue:\\

\noindent \texttt{{\sc fellwalker.allowedge} = 0 \\
{\sc fellwalker.cleaniter} = 5 \\
{\sc fellwalker.fwhmbeam} = 1 \\
{\sc fellwalker.minpix} = 12 \\
{\sc fellwalker.mindip} = 1.5 \\
{\sc fellwalker.maxjump} = 3 \\
{\sc fellwalker.minheight} = 3 \\
{\sc fellwalker.noise} = 1 \\
{\sc fellwalker.shape} = ellipse\\
}

\clearpage
\onecolumn

\noindent
Author affiliations:\\
$^{1}$Astrophysics Research Institute, Liverpool John Moores University, IC2, Liverpool Science Park, 146 Brownlow Hill, Liverpool, L3 5RF, UK\\
$^{2}$Department of Physics and Astronomy, University of Calgary, 2500 University Drive NW, Calgary, Alberta T2N 1N4, Canada\\
$^{3}$School of Physical Sciences, Ingram Building, University of Kent, Canterbury, Kent CT2 7NH, UK\\
$^{4}$Centre for Astrophysics Research, Science \& Technology Research Institute, University of Hertfordshire, College Lane, Hatfield, Herts AL10 9AB, UK\\
$^{5}$East Asian Observatory, 660 N. Aohoku Place, University Park, Hilo, Hawaii 96720, USA\\
$^{6}$School of Physics and Astronomy, Cardiff University, Cardiff CF24 3AA, UK\\
$^{7}$Astrophysics Group, Cavendish Laboratory, J J Thomson Avenue, Cambridge CB3 0HE, UK\\
$^{8}$Kavil Institute for Cosmology, Institute of Astronomy, University of Cambridge, Madingley Road, Cambridge CB3 0HA, UK\\
$^{9}$Astrophysics Group, School of Physics, University of Exeter, Stocker Road, Exeter EX4 4QL, UK\\
$^{10}$Department of Physics and Astronomy, James Madison University, MSC 4502, 901 Carrier Drive, Harrisonburg, VA 22807, USA\\
$^{11}$SKA Headquarters, University of Manchester, Manchester M13 9PL, UK\\
$^{12}$Department of Physics and Astronomy, University of Waterloo, Waterloo, ON N2L 3G1, Canada\\
$^{13}$Department of Physics and Astronomy, University of British Columbia, 6224 Agricultural Road, Vancouver, BC V6T 1Z1, Canada\\
$^{14}$School of Physics and Astronomy, University of Leeds, Leeds LS2 9JT, UK\\
$^{15}$501 Large Synoptic Survey Telescope, 950 N. Cherry Ave, Tuscon, Arizona 835719, USA\\
$^{16}$ Max-Planck Institute for Astronomy, K\"{o}nigstuhl 17, 619117 Heidelberg, Germany\\
$^{17}$INAF-IAPS, via del Fosso del Cavalier\'{e} 100, 00133 Roma, Italy\\
$^{18}$Departamento de Fisica, Universidad de Atacama, Copayapu 485, Copiap\'{o}, Chile\\
$^{19}$National Research Council of Canada, 5071 West Saanich Road, Victoria, BC V9E 2E7, Canada\\
$^{20}$Astrophysics Group, Keele University, Keele, Staffordshire ST5 5BG, UK\\
$^{21}$Jodrell Bank Centre for Astrophysics, School of Physics and Astronomy, The University of Manchester, Oxford Road, Manchester M13 9PL, UK\\
$^{22}$Department of Physics and Astronomy, University of Victoria, Victoria, BC V8P 1A1, Canada\\
$^{23}$Centre de Recherche en Astrophysique du Qu\'{e}bec, D\'{e}partement de physique, de g\'{e}nie physique et d'optique, Universit\'{e} Laval, QC G1K 7P4, Canada\\
$^{24}$Canadian Institute for Theoretical Astrophysics, University of Toronto, 60 St George Street, Toronto, ON M5S 3H8, Canada\\
$^{25}$Department of Physics and Astronomy, The Open University, Walton Hall, Milton Keynes, MK7 6AA, UK\\
$^{26}$RAL Space, STFC Rutherford Appleton Laboratory, Chilton, Didcot, Oxfordshire OX11 0QX, UK\\
$^{27}$National Astronomical Observatories, Chinese Academy of Sciences, 20A Datun Road, Chaoyang District, Beijing, 100012, China\\

\bsp
\label{lastpage}

\end{document}